\documentclass[usenatbib]{mn2e}
\usepackage{graphicx}
\usepackage{ifthen}
\usepackage{amsmath}
\usepackage{url}
\usepackage{rotating}
\usepackage{xcolor}

\def\ltsima{$\; \buildrel < \over \sim \;$}
\def\lta{\lower.5ex\hbox{\ltsima}}
\def\gtsima{$\; \buildrel > \over \sim \;$}
\def\simgt{\lower.5ex\hbox{\gtsima}}
\def\kms{{\rm\,km \; s^{-1}}}

\def\AA{$\; \buildrel \circ \over {\rm A}$}

\def\s{\ifmmode \widetilde \else \~\fi}
\def\={\overline}

\def\spose#1{\hbox to 0pt{#1\hss}}

\def\lta{\mathrel{\spose{\lower 3pt\hbox{$\mathchar"218$}}
     \raise 2.0pt\hbox{$\mathchar"13C$}}}
\def\gta{\mathrel{\spose{\lower 3pt\hbox{$\mathchar"218$}}
     \raise 2.0pt\hbox{$\mathchar"13E$}}}
\def\Dt{\spose{\raise 1.5ex\hbox{\hskip3pt$\mathchar"201$}}}    % upper case
\def\dt{\spose{\raise 1.0ex\hbox{\hskip2pt$\mathchar"201$}}}    % lower case

\def\dotsfill{\leaders\hbox to 1em{\hss.\hss}\hfill}

\loadboldmathitalic 
\title[The Pristine Dwarf-Galaxy survey IV]{The Pristine Dwarf-Galaxy survey - IV. Probing the outskirts of the dwarf galaxy Bo\"otes~I}
\author[N. Longeard et al.] {Nicolas Longeard$^{1}$, Pascale Jablonka$^{1,2}$, Anke Arentsen$^{3}$, Guillaume F. Thomas$^{4,5}$, 
\newauthor David S. Aguado$^{6}$, Raymond G. Carlberg$^{7}$, Romain Lucchesi$^{1}$, Khyati Malhan$^{8}$,  
\newauthor Nicolas Martin$^{3,9}$, Alan W. McConnachie$^{10}$, Julio F. Navarro$^{11}$, Rub\'en S\'anchez-Janssen$^{12}$,
\newauthor Federico Sestito$^{13}$, Else Starkenburg$^{14,15}$, Zhen Yuan$^{3}$ \\
$^{1}$ Laboratoire d'astrophysique, \'Ecole Polytechnique F\'ed\'erale de Lausanne (EPFL), Observatoire, 1290 Versoix, Switzerland\\
$^{2}$ GEPI, Observatoire de Paris, Universit\'e PSL, CNRS, Place Jules Janssen, F-92195 Meudon, France\\
$^{3}$ Universit\'e de Strasbourg, CNRS, Observatoire astronomique de Strasbourg, UMR 7550, F-67000 Strasbourg, France\\
$^{4}$ Instituto de Astrof\'isica de Canarias, V\'ia L\'actea, 38205 La Laguna, Tenerife, Spain \\
$^{5}$ Universidad de La Laguna, Departamento de Astrof\'isica, 38206 La Laguna, Tenerife, Spain \\
$^{6}$ Institute of Astronomy, University of Cambridge, Madingley Road, Cambridge CB3 0HA, UK \\
$^{7}$ Department of Astronomy \& Astrophysics, University of Toronto, Toronto, ON M5S 3H4, Canada \\
$^{8}$ The Oskar Klein Centre for Cosmoparticle Physics, De- partment of Physics, Stockholm University, AlbaNova, 10691 Stockholm, Sweden \\
$^{9}$ Max-Planck-Institut f\"ur Astronomy, K\"onigstuhl 17, D-69117, Heidelberg, Germany\\
$^{10}$ NRC Herzberg Astronomy and Astrophysics, 5071 West Saanich Road, Victoria, BC V9E 2E7, Canada\\
$^{11}$ University of Victoria, 3800 Finnerty Rd, Victoria, BC, V8P 5C2, Canada \\
$^{12}$ STFC UK Astronomy Technology Centre, Royal Observatory, Blackford Hill, Edinburgh, EH9 3HJ, UK \\
$^{13}$ Department of Physics and Astronomy, University of Victoria, PO Box 3055, STN CSC, Victoria, BC V8W 3P6, Canada \\
$^{14}$ Leibniz Institute for Astrophysics Potsdam (AIP), An der Sternwarte 16, 14482 Potsdam, Germany\\
$^{15}$ Kapteyn Astronomical Institute, University of Groningen, Landleven 12, 9747 AD Groningen, The Netherlands\\
}

\date{\today}
\begin{document} 
\maketitle 
\begin{abstract}
We present a new spectroscopic study of the dwarf galaxy Bootes~I (Boo~I) with data from the Anglo-Australian Telescope and its AAOmega spectrograph together with the Two Degree Field multi-object system. We observed 36 high-probability Boo~I stars selected using Gaia Early Data Release 3 proper motions and photometric metallicities from the Pristine survey. Out of those, 27 are found to be Boo~I's stars, resulting in an excellent success rate of $75$\% at finding new members. Our analysis uses a new pipeline developed to estimate radial velocities and equivalent widths of the calcium triplet lines from Gaussian and Voigt line profile fits. The metallicities of 16 members are derived, including 3 extremely metal-poor stars ([Fe/H] $< -3.0$), which translates into a success rate of 25\% at finding them with the combination of Pristine and Gaia. Using the large spatial extent of our new members that spans up to 4.1 half-light radii and spectroscopy from the literature, we find a systemic velocity gradient of $0.40 \pm 0.10$ km s$^{-1}$ arcmin$^{-1}$ and a small but resolved metallicity gradient of $-0.008 \pm 0.003$ dex arcmin$^{-1}$. Finally, we show that Boo~I is more elongated than previously thought with an ellipticity of $\epsilon = 0.68 \pm 0.15$. Its velocity and metallicity gradients as well as its elongation suggest that Boo~I may have been affected by tides, a result supported by direct dynamical modelling.
\end{abstract}
 

\section{Introduction}

The $\Lambda$CDM cosmological model predicts the existence of low-mass, low-luminosity galaxies orbiting around massive host galaxies such as our Milky Way (MW) (\citealt{bullock01}, \citealt{benson02}, \citealt{wechsler09}, \citealt{bullock17}). These satellites are also supposed to be the most dark matter (DM) dominated structures in the Universe. The first dwarf galaxies discovered almost a century ago (Sculptor and Fornax, \citealt{shapley38b}) are bright and massive objects. Together with 6 other discoveries which followed over the decades, they constitute the so-called ``classical dwarf spheroidal'' (dSph) satellites of the MW. 

More recently, large photometric surveys such as the Sloan Digital Sky Survey \citep[SDSS]{york00}, the Panoramic Survey Telescope and Rapid Response System $3\Pi$ \citep[PanSTARRS $3\Pi$]{chambers16} and the Dark Energy Survey \citep[DES]{des05} have enabled the discovery of dozens of smaller and smaller systems that are now commonly referred to as Ultra-Faint Galaxies (UFDs). These discoveries were accompanied by efforts to spectroscopically follow-up all systems that were discovered (e.g. \citealt{kleyna05}, \citealt{munoz06}, \citealt{martin07}, \citealt{simon_geha07}, \citealt{koposov11}, \citealt{kirby13a}, \citealt{martin16_dra}, \citealt{longeard18}, \citealt{longeard20}). These spectroscopic observations have two main focus points, namely to derive their kinematics and their metallicity properties. In particular, the kinematics of UFD candidates are crucial as their velocity dispersion can be directly linked to their DM halo mass, should they have one \citep[W10]{wolf10}. The DM content of all these UFDs enables direct comparison with cosmological simulations (\citealt{springel08}, \citealt{vogelsberger14}, \citealt{fattahi16}, \citealt{sawala16}, \citealt{read19}). However, two major assumptions are necessary to link kinematics and DM mass in the model of W10 that is widely used for UFDs: (1) the system must be in dynamical equilibrium and (2) its velocity dispersion profile must be flat, i.e. independent of the distance to the centre of the UFD.

Classical dSphs are extended with a size of several hundreds of pc (\citealt{mcconnachie12}, \citealt{simon19}) and are densely populated systems, but UFDs are much fainter and their contrast on the sky with respect to MW halo stars can be even lower. As a result, it is challenging to find members in UFDs with spectroscopy. In addition, UFDs have a typical size of  tens of pc which is larger than the field-of-view (FoV) of most spectrographs at their distances. As a result, the vast majority of spectroscopic campaigns targeting UFDs have been focused on their inner region (roughly within $1$ half-light radius). These observations have been very successful at identifying a large number of member stars in these elusive systems, providing radial velocities and metallicities. They also have led to the detection of 54 extremely metal-poor (EMP) stars ([Fe/H] $< -3.0$) in UFDs according to the Stellar Abundances for Galactic Archaeology Database (SAGA) database\footnote{http://sagadatabase.jp}. Such stars provide valuable information about their early formation and evolution (\citealt{tolstoy09}, \citealt{frebel_norris15}) .

On the other hand, because these observations have largely targeted the inner regions of UFDs, they provide very little information on the properties of these satellites in their outer regions where departures from dynamical equilibrium and the effects of tides would be more clearly noticeable. Although such ventures into the outer regions have been attempted before (e.g. \citealt{koposov11}, \citealt{fritz19}, \citealt{chiti21}, \citealt{longeard21}), they are still limited by the observational constrains mentioned above and generally have a very low success rate in confirming members. It is therefore currently challenging to assess the reliability of the overall DM mass of UFDs known. Are all of them reliable, and if not, how many deviates from dynamical equilibrium and have a more complex dynamics and/or metallicity properties ? \\

To shed light on these two important questions, one needs to probe the outskirts of UFDs. Their stars tend to be very difficult to identify, as they are drowned by stars in the MW halo. Such studies must therefore adopt additional tools that exploit the kinematic and metallicity features that distinguish UFD stars from the MW stellar halo. Proper motions (PM) from the Gaia mission (\citealt{brown18}, \citealt{gaia21}) are one of those tools, helping to identify UFD members using the systematic motion of the satellite. In addition to Gaia, our team benefits from the photometric metallicities of the Pristine survey (Starkenburg et al. 2017). The Pristine metallicities are reliable to very low metallicities, and are particularly well-suited to detect the metal-poor population typical of UFDs. \citet{longeard21} show that using Pristine alone can improve the success rate of finding new member stars by a factor of 3, from 20\% using only broad-band photometric constraints to 60\% by adding Pristine. Together, these datasets allow us to hunt for UFD member stars at large distances and to more efficiently detect EMP stars (\citealt{youakim17}, \citealt{aguado19}).

In this context, we present here a new set of spectroscopic observations of the faint UFD Boötes~I (Boo~I). This exceptionally large FoV combined with the Gaia Data Release 3 and the Pristine survey enable us to probe the outskirts of Boo~I up to a distance of 4.1 half-light radii (r$_h$). Boo~I was discovered by \citet{belokurov06} and is the most extensively spectroscopically-studied UFD (\citealt{martin07}, \citealt{norris08}, \citealt{norris10}, \citealt{koposov11}, \citealt{jenkins21} for studies identifying new Boo~I members and not re-analyzing known Boo~I stars at high resolution). These studies have focused on the central region of the system, with $\sim$ 80\% of the known members enclosed within 1r$_h$ and 94\% within 2r$_h$ to maximize the chance of finding member stars, since there should be $50$\% of stars within 1r$_h$ and $85$\% within 2r$_h$ when an exponential radial profile is considered. Metallicity measurements for the stars beyond 1r$_h$ only represent about $17$\% of all Boo~I members with a known [Fe/H]. Therefore, Boo~I is a good target for such a study, even more so since the spectroscopy presented by \citet{koposov11} suggest that the kinematics of the system might be more complex that what is commonly assumed for UFDs with the possible existence of two dynamically different population in the satellite. However, since the kinematics of Boo~I can also be reasonably well described by a single population, these results are not definitive.

This work first details the new spectroscopic observations (section 2.1), as well as the new pipeline developed by our team to derive radial velocities and metallicities from stellar spectra around the calcium triplet (CaT) region (section 2.2). We study the kinematics and metallicity properties of Boo~I in its outskirts and discuss whether its dynamics (section 3.1) and metallicity properties (section 3.2) exhibit spatial gradients that may reflect departure from equilibrium or the possibility that Boo I may have been affected by tidal interactions with the MW (section 3.3). We conclude with a discussion in section 4.

\section{Spectroscopic observations}

This section details the selection, observations and reduction of the data as well as our pipeline to derive radial velocities and equivalent widths from the spectra.

\subsection{Data selection and acquisition}
Our target selection relies on data from the Pristine survey \citep{starkenburg17}. Pristine is a photometric survey based on a narrow-band, metallicity-sensitive photometry centered on the calcium H\&K doublet and has proven to be efficient at identifying metal-poor stars such as those characteristic of the UFDs stellar population (\citealt{youakim17}, \citealt{aguado19}, \citealt{arentsen20}, \citealt{longeard21}). Therefore, we have reliable photometric metallicity estimates for Boo~I down to $g_0^{\mathrm{SDSS}} \sim 21.5$. To build our spectroscopic target list, the following criteria are used:

\begin{itemize}
\item Using the photometry from the SDSS and the Darmouth isochrone \citep{dotter08} fitted to the Boo~I stellar population ($A = 13$ Gyr, [Fe/H] $= -2.3$, [$\alpha$/Fe] $= 0.0$, $m - M = 19.11$), we remove any star located further than 0.15 mag from the isochrone in the (($g-i$)$_0$,$i_0$) colour-magnitude diagram (CMD).
\item Any star with a Pristine photometric metallicity above $-1.0$ dex is also discarded. 

\item When available, the PMs from the Gaia Early DR3 \citep[eDR3]{gaia21} is also used to keep only stars with a PM compatible within $2\sigma$ from that of Boo~I (\citealt{mcconnachie20}, \citealt{battaglia22}, $\langle \mu^{*}_{\alpha} \rangle = -0.39 \pm 0.01$ mas yr$^{-1}$ and $\langle \mu_{\delta} \rangle = -1.06 \pm 0.01$ mas yr$^{-1}$ ).

\end{itemize}

\begin{figure*}
\begin{center}
\centerline{\includegraphics[width=\hsize]{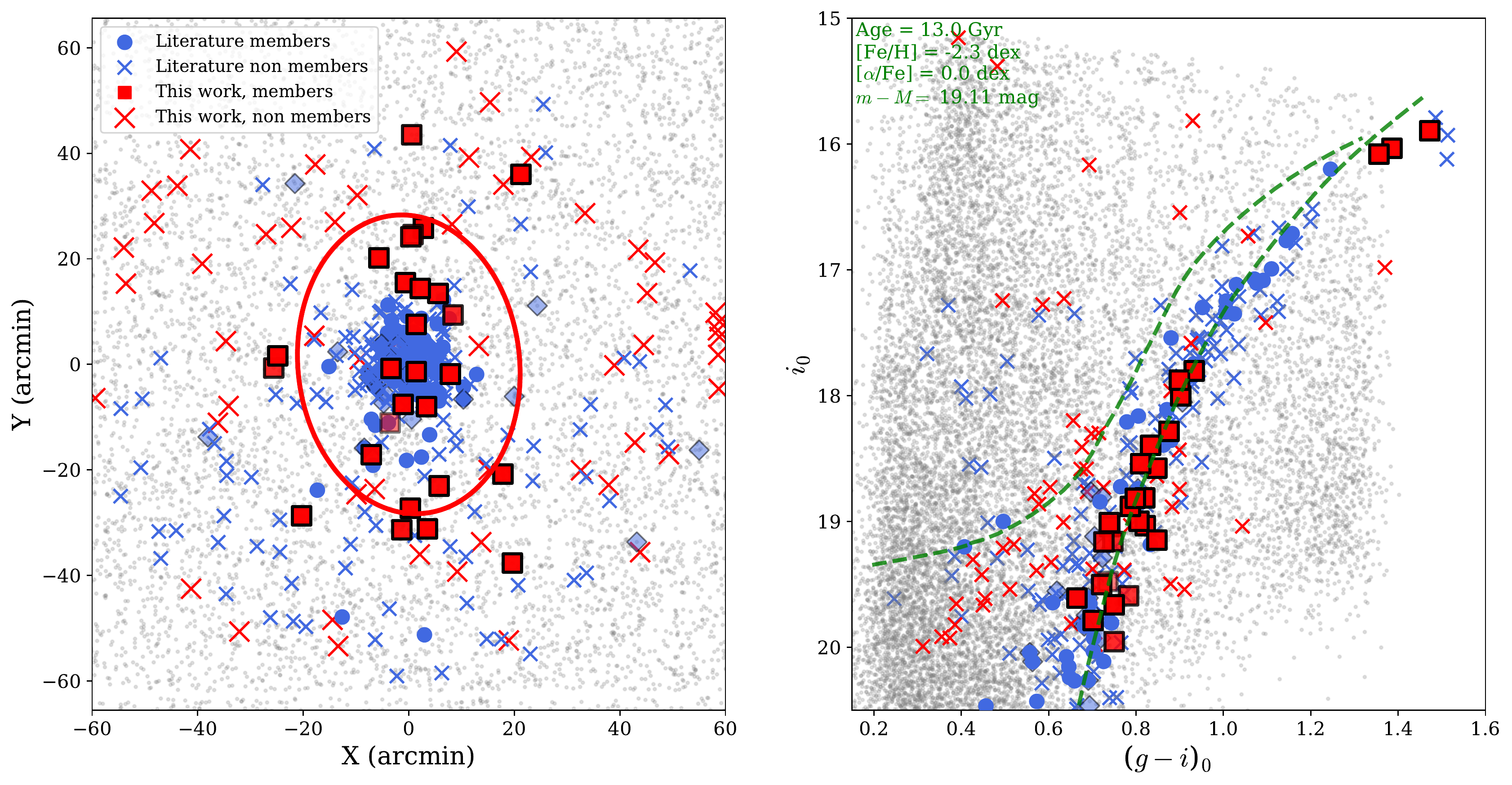}}
\caption{\textit{Left panel:} Spatial overview of the AAT spectroscopic sample. Newly discovered members are shown as red squares and their opacity depends on their dynamical membership probability. In particular, the  AAT member represented with a significantly transparent symbol is the only one with a membership probability below $80$\%, with $P_\mathrm{mem} \sim 60$\%. Non-members from the AAT sample are shown as red crosses. Previously known members from the literature are represented as smaller blue circles. Members from the literature that are not members according to this work are shown as transparent blue diamonds. The two half-light radii of Boo~I as inferred by \citet[M18]{munoz18} are shown as a red ellipse. \textit{Right panel:} CMD  of our spectroscopic sample superimposed with a metal-poor Darmouth isochrone at the distance of Boo~I.}
\label{field} 
\end{center}
\end{figure*}

These seemingly generous constraints, especially the photometric metallicity one given the systemic metallicity of the system ([Fe/H]$_\mathrm{BooI}$ $\sim -2.35$ dex), are enabled by the large number of fibers available, which far exceeds our number of promising targets and allows us to cast a wide net. Finally, no stars previously observed in the literature at the time were targeted by these observations. In total, the final target list was composed of $92$ stars including 36 probable Boo~I Red Giant Branch (RGB) members, 18 potential Horizontal Branch (HB) stars and $12$ potentially very metal -poor halo stars. The rest are low probability potential Boo~I members that do not satisfy CMD or PM criteria of the 36 high probability stars but that were observed because of the large number of fibers. An overview of the spatial and CMD locations of these stars can be found in Figure \ref{field}.

The data were obtained as a filler program for the Pristine Inner Galaxy Survey \citep[PIGS]{arentsen20}, at the beginnings of the nights on the 16/06/2020, 17/06/2020 and 19/06/2020 with two sub-exposures taken each night, on the Anglo-Australian Telescope \citep[AAT]{saunders04} and its AAOmega spectrograph (\citealt{lewis02}, \citealt{sharp06}) together with the Two Degree Field (2dF) multi-object system \citep{cannon97}. This setup benefits from $\sim 360$ science fibers and $\sim 40$ fibers for sky spectra and guiding. Only one two-degree field was needed to observe all our targets. The total exposure were divided into 6 sub-exposures of 2700 seconds each for a total of 4.5 hours. The gratings used were 580V for low-resolution spectra in the optical (R $\sim$ 1300, 3700-5500 A), and 1700D for calcium triplet spectra with a spectral resolution R of $\sim 11000$. Only the red part of the spectra (from $840$ to $880$ nm) is used for the rest of this work.

\subsection{Reduction and pipeline}

\begin{figure*}
\begin{center}
\centerline{\includegraphics[width=\hsize]{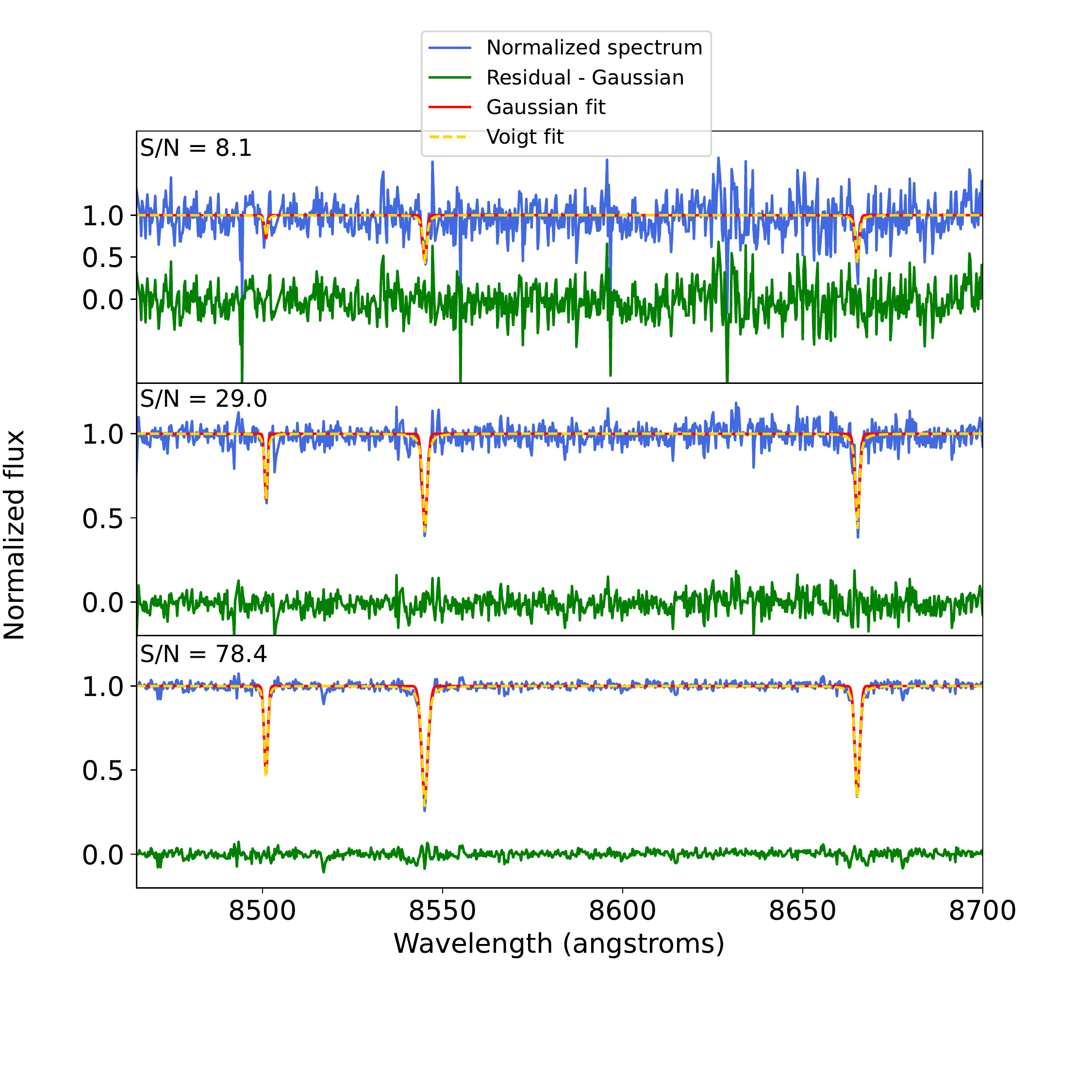}}
\caption{Spectra of three member stars in our AAT dataset centred on the calcium triplet lines. Each star represents respectively the low, mid and high S/N regimes. The normalised spectra are shown with solid blue lines while the fits derived from our pipeline for Gaussian and Voigt line profiles, detailed in section 2.2, are shown with dashed red and yellow lines respectively. For clarity, only the residuals in the Gaussian cases are shown (green). These stars have a radial velocity of $102.2 \pm 2.8$, $103.0 \pm 1.0$ and $102.2 \pm 0.7$ km s$^{-1}$ from top to bottom.}
\label{spectrum} 
\end{center}
\end{figure*}

The resulting spectra were reduced using the AAT 2DFDR\footnote{https://aat.anu.edu.au/science/software/2dfdr} package and the standard settings, with two small exceptions. Details on the data reduction can be found in the ``Data Reduction'' section of \citet{arentsen20}. Three examples of spectra for low (8.1), mid (29.0) and high (78.4) signal-to-noise (S/N) ratios are shown in Figure \ref{spectrum}. The S/N is computed in the CaT region.

\subsubsection{Technical description of the analysis pipeline}

To derive the radial velocities and equivalent widths (EWs) for the CaT lines, we developed a new pipeline that finds the continuum, performs sky-subtraction, removal of telluric lines, and fits the calcium triplet lines with both Gaussian and Voigt profiles. First, we use the method of \citet{battaglia08} to find the best continuum and normalize the spectra through an iterative k-sigma clipping, non-linear filter. We get rid of potential cosmic rays by discarding any emitted flux value 6$\sigma$ above the continuum, with $\sigma$ being the flux uncertainty of a given spectrum. The value of $\sigma$ is obtained by computing the standard deviation of the continuum around the CaT lines excluding all spectral features. Sky-subtraction and removal of telluric lines are performed by minimizing the difference between a stellar spectrum and all the sky spectra available at the CaT lines region. Since the sky and stellar spectra can have very different fluxes, the minimisation is performed by scaling the sky spectrum with a multiplying factor varying between 0.1 and 10. The best sky spectrum is then subtracted from the stellar one and this operation is repeated a second time to account for the fact that no unique sky spectrum will perfectly match the sky features present in a given stellar spectrum. 

At this stage, we have normalised, cosmic rays-cleaned, sky-substracted and telluric-corrected spectra. Once these steps are finished, the pipeline derives the relevant scientific quantities, i.e. the radial velocity and EWs. The first step is to have a first initial guess of the radial velocity of the star. The pipeline first derives a smoothed spectrum for each star with a gaussian kernel of width corresponding to 4 elements of resolution in order to highlight the CaT lines. We then compute the cross-correlation of these smoothed spectra with a simulated spectrum containing only the CaT lines modelled with fixed Gaussian profiles. Each Gaussian profile is defined as the following:

\begin{equation}
\begin{aligned}
\mathcal{G}_\mathrm{line} = \frac{1}{\sigma_\mathrm{line}\sqrt{2 \pi}} \mathrm{exp(}-0.5\frac{(\lambda_0- \Delta\lambda(\mathrm{v}))^{2}}{\sigma_\mathrm{line}^{2}}    \mathrm{)}
\end{aligned}
\end{equation}

\noindent with $\sigma_\mathrm{line}$ being the standard deviation, $\lambda_0$ the theoretical location of each line at rest and $\Delta\lambda(\mathrm{v})$ the doppler shift directly linked to the radial velocity $\mathrm{v}$. The Voigt profiles are defined with the approximation described by \citet{mclean94} that takes three parameters into account: the amplitude of the Lorentzian component of the Voigt profile and the standard deviations of both the Lorentzian and Gaussian components. 

In this step, only the radial velocity varies. The maximum of this correlation yields the initial radial velocity guess. The derivation of the final scientific parameters will be performed around the initial guess. Using the simulated spectrum with only the CaT lines as spectral features shifted at the initial radial velocity guess, we proceed to derive the radial velocity and EWs by fitting the spectra with a Monte Carlo Markov Chain algorithm \citep[MCMC]{hastings70}. The central wavelengths (and therefore the radial velocities, $\mathrm{v}$), the normalised fluxes of each line $a_1$, $a_2$ and $a_3$ as well as their standard deviations are fitted by minimising the following likelihood:

\begin{equation}
\mathcal{L}_k = \frac{1}{\sigma_k\sqrt{2 \pi}} \mathrm{exp(}-0.5\frac{(y_{obs,k} - y_{s,k})^{2}}{\sigma_k^{2}}    \mathrm{)}
\end{equation}

\noindent where $\sigma_k$ stands for the flux uncertainty for the $k$-th star, y$_{obs,k}$ is the observed stellar spectrum and y$_{s,k}$ is the simulated one. The three lines are fitted simultaneously and their shapes are constrained with respect to those of the two others. The second line is constrained to be deeper than the third one which is set to be deeper than the first one. The first line also cannot be narrower than the other two. Similarly, the third line is set to be narrower than the second one.

For each star, the MCMC is ran for a million iterations and the results are chosen to be the parameters that maximise the likelihood $\mathcal{L}_k$. To obtain the EWs, each calcium triplet line from the best simulated spectrum is integrated within a 15\AA $\;$ window. Finally, in the Gaussian profiles case, the EWs are divided by a factor of 1.1 to account for the poor modelling of the wings of the lines by gaussian line profiles \citep{battaglia08}.

\subsubsection{Testing the pipeline's performance}

\begin{figure}
\begin{center}
\centerline{\includegraphics[width=\hsize]{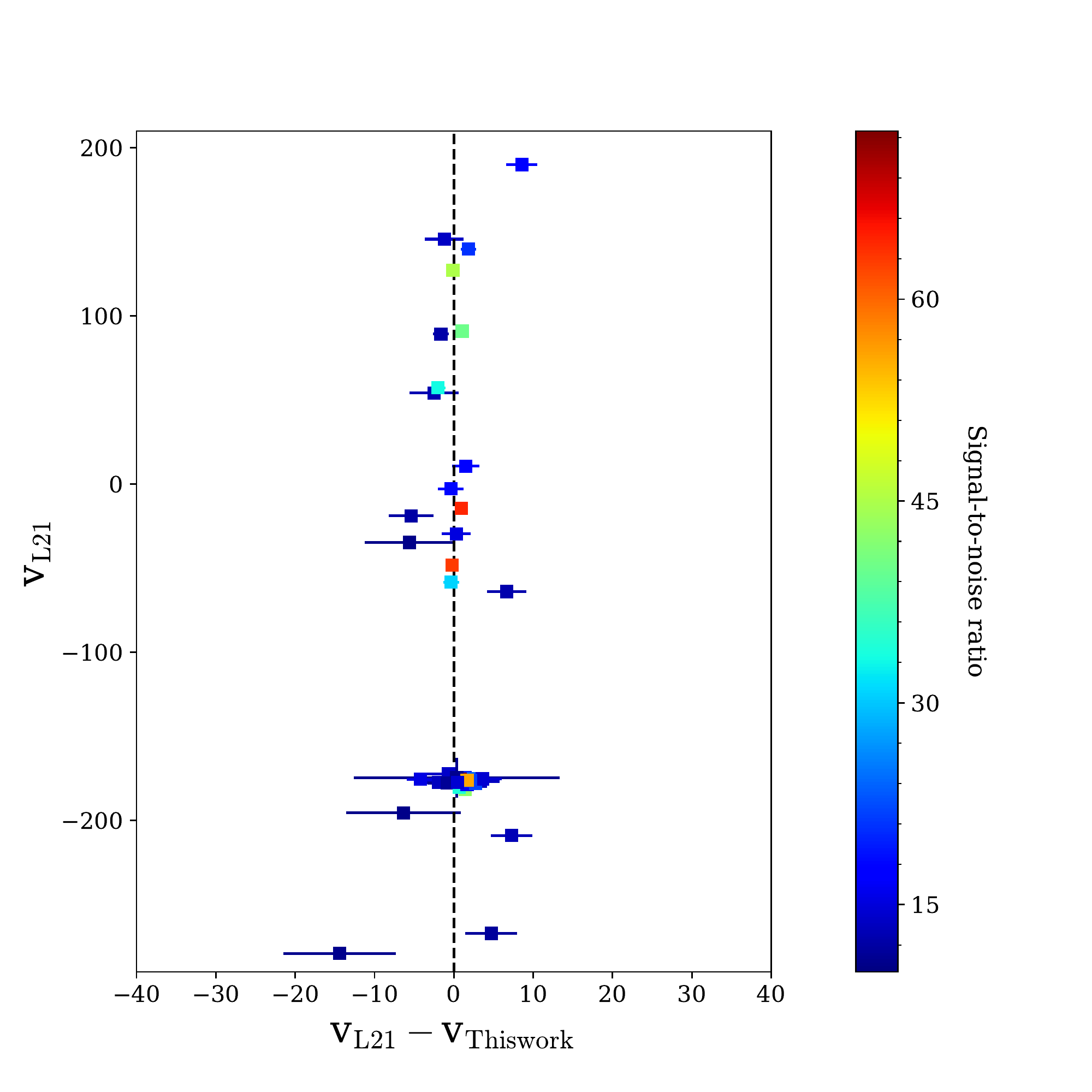}}
\caption{Difference between the radial velocities obtained with our new pipeline and the ones from L21 on the x-axis with respect the L21 velocities on the y-axis. The data are colour-coded according to their S/N ratios and the black dashed line shows the 0 km s$^{-1}$ offset line.}
\label{me_vs_L21} 
\end{center}
\end{figure}

\begin{figure}
\begin{center}
\centerline{\includegraphics[width=\hsize]{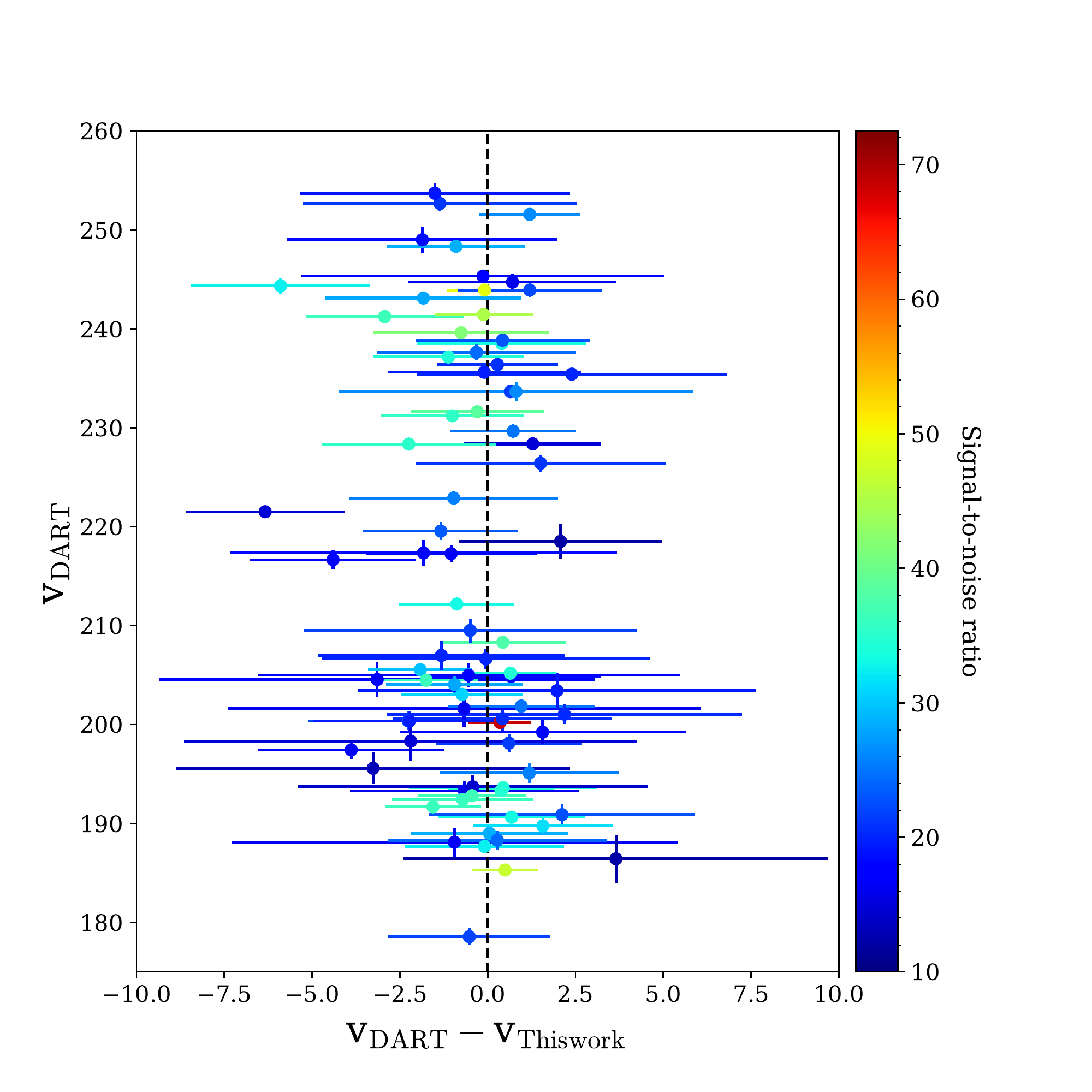}}
\caption{Difference between the radial velocities obtained with our new pipeline and the ones from DART on the x-axis with respect the DART velocities on the y-axis for a sample of 86 Sextans stars. The data are colour-coded according to their S/N ratios and the black dashed line shows the 0 km s$^{-1}$ offset line.}
\label{me_vs_dart_v} 
\end{center}
\end{figure}

\begin{figure}
\begin{center}
\centerline{\includegraphics[width=\hsize]{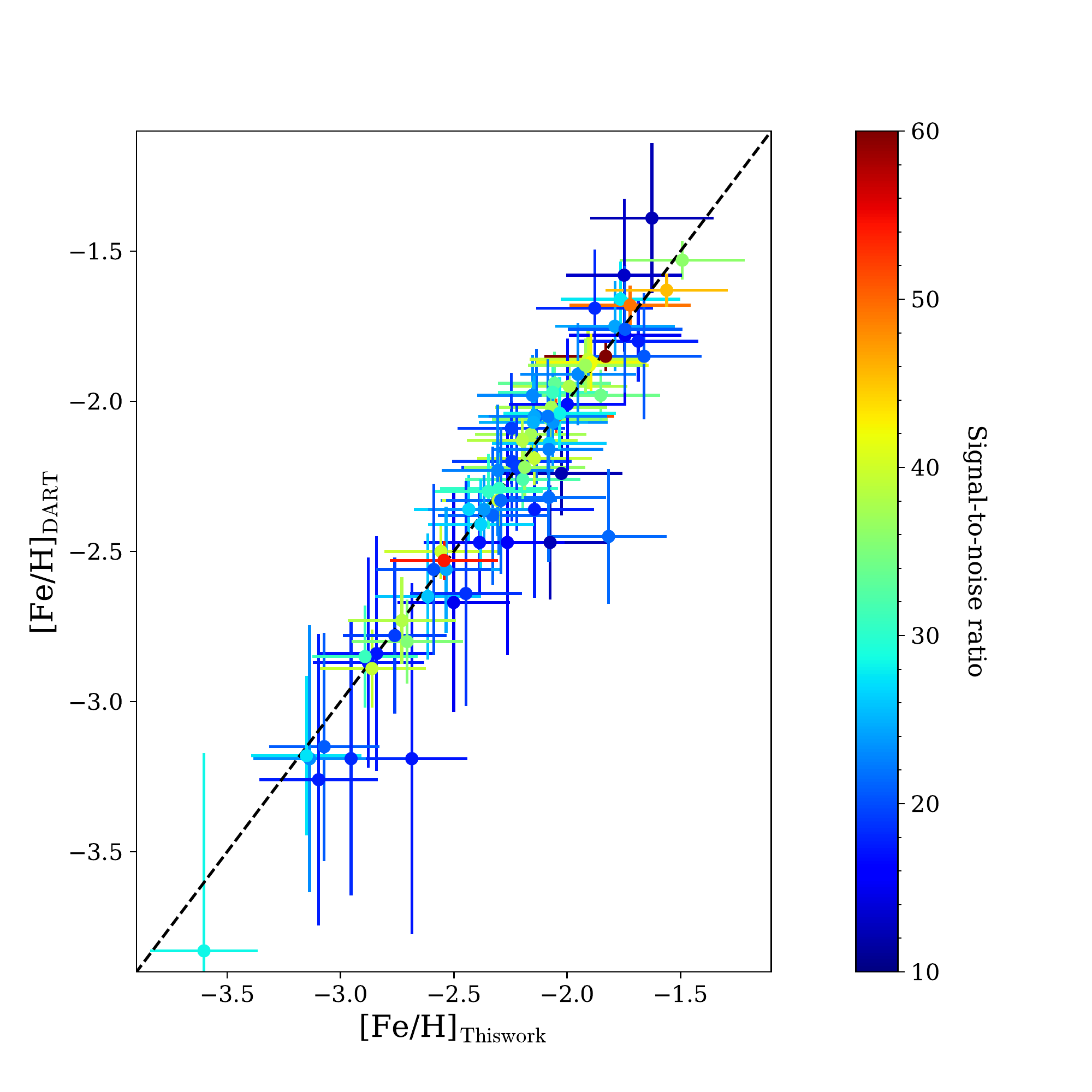}}
\caption{Comparison between the metallicities obtained with our new pipeline on the x-axis with the ones from the DART survey on the y-axis for a sample of 86 Sextans stars. The data are colour-coded according to their S/N ratios and the black dashed line shows the 1:1 line.}
\label{me_vs_dart} 
\end{center}
\end{figure}

\begin{figure}
\begin{center}
\centerline{\includegraphics[width=\hsize]{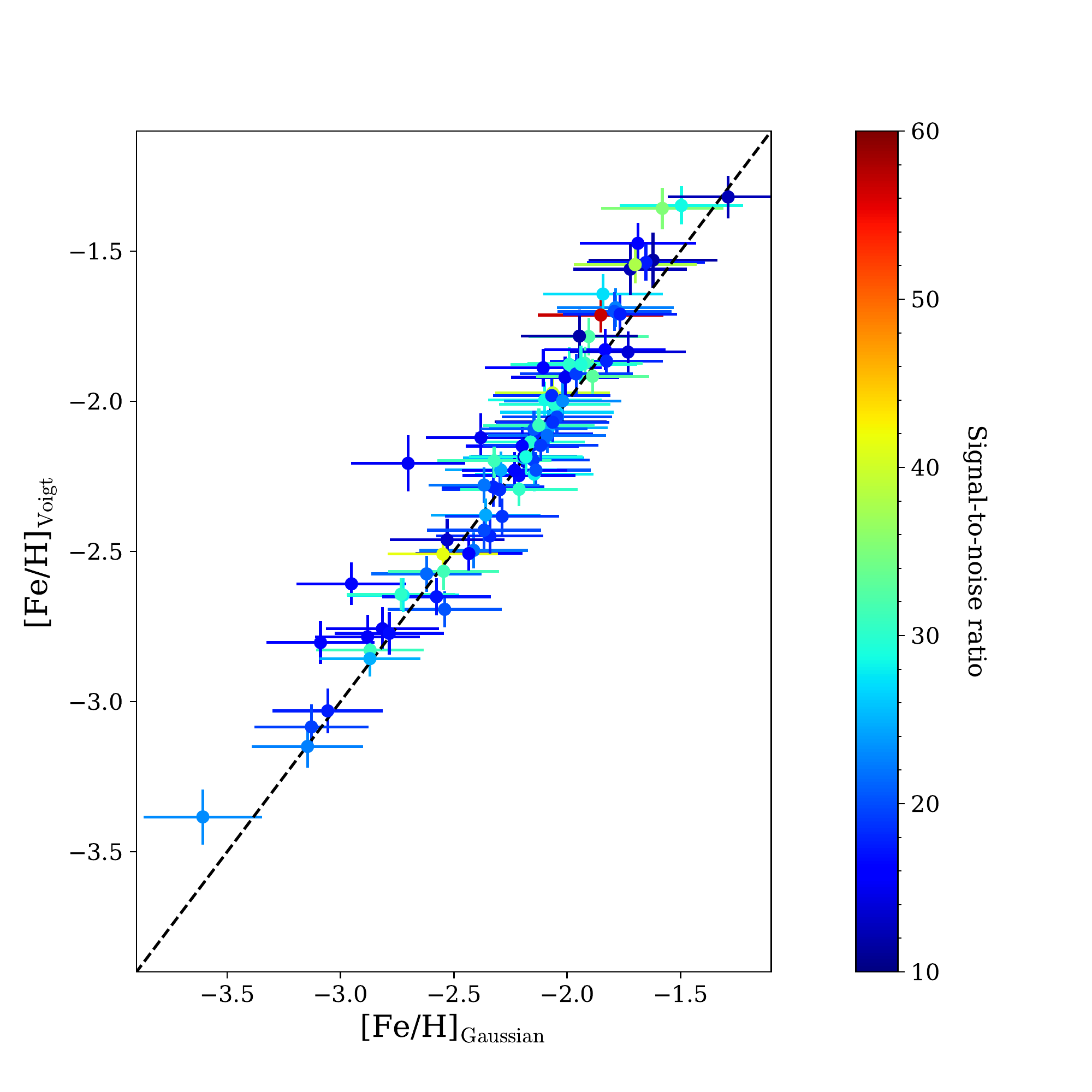}}
\caption{Comparison between the metallicities obtained with our new pipeline with Gaussian line profiles on the x-axis and with Voigt profiles on the y-axis for the sample of Sextans stars. The data are colour-coded according to their S/N ratios and the black dashed line shows the 1:1 line.}
\label{G_vs_V_dart} 
\end{center}
\end{figure}

The pipeline is then thoroughly tested against previous spectroscopic studies to ensure that both the kinematics and the EWs (and therefore metallicities) are properly derived. 

We take as reference 86 stars from Sextans and analysed by the \textit{Dwarf galaxy Abundances and Radial-velocities Team} (\citealt[DART]{tolstoy06}, \citealt{battaglia11}). Choosing DART is ideal as their results use the same method as our new pipeline to derive metallicities (see section 3.2), i.e. fitting the spectra with Gaussian profiles and using the empirical calibration of \citet{starkenburg10} to translate the CaT EWs into [Fe/H]. \citet{battaglia11} also use a setup with a resolution not too far from the one in this work (R$_\mathrm{DART} \sim 8000$). Since this subsample only contains stars with S/N $> 10$, we supplement the reference sample with stars from \citet{longeard21} and the Sagittarius~2 globular cluster containing metal-poor stars down to a S/N of 3 to compare radial velocities in the low S/N regime. The results of this comparison are shown in Figure \ref{me_vs_L21} and \ref{me_vs_dart_v} for velocities and Figure \ref{me_vs_dart} for metallicities. They both show that our pipeline provides excellent radial velocities and metallicity measurements for all S/N regimes. By assuming that the difference between our new pipeline and DART results are normally distributed for velocities and metallicities, we find a negligible velocity bias of $0.02 \pm 0.4$ km s$^{-1}$ and a scatter below $1.0$ km s$^{-1}$ at the 95\% confidence limit. For the metallicity, the bias, also negligible, is $0.03 \pm 0.04$ and the scatter below $0.09$ at the 95\% confidence limit. Figure \ref{G_vs_V_dart} also shows that metallicities derived with Gaussian and Voigt line profiles are in agreement for this sample of Sextans stars with a mean difference of $0.02 \pm 0.03$ and a scatter below $0.08$ at the 95\% confidence limit. These comparisons show that there is no statistically significant bias in metallicity or velocity.

\section{Results}

We present in this section the results of our spectroscopic analysis, both dynamical and in terms of metallicity.

\subsection{Dynamical analysis}

\begin{figure}
\begin{center}
\centerline{\includegraphics[width=\hsize]{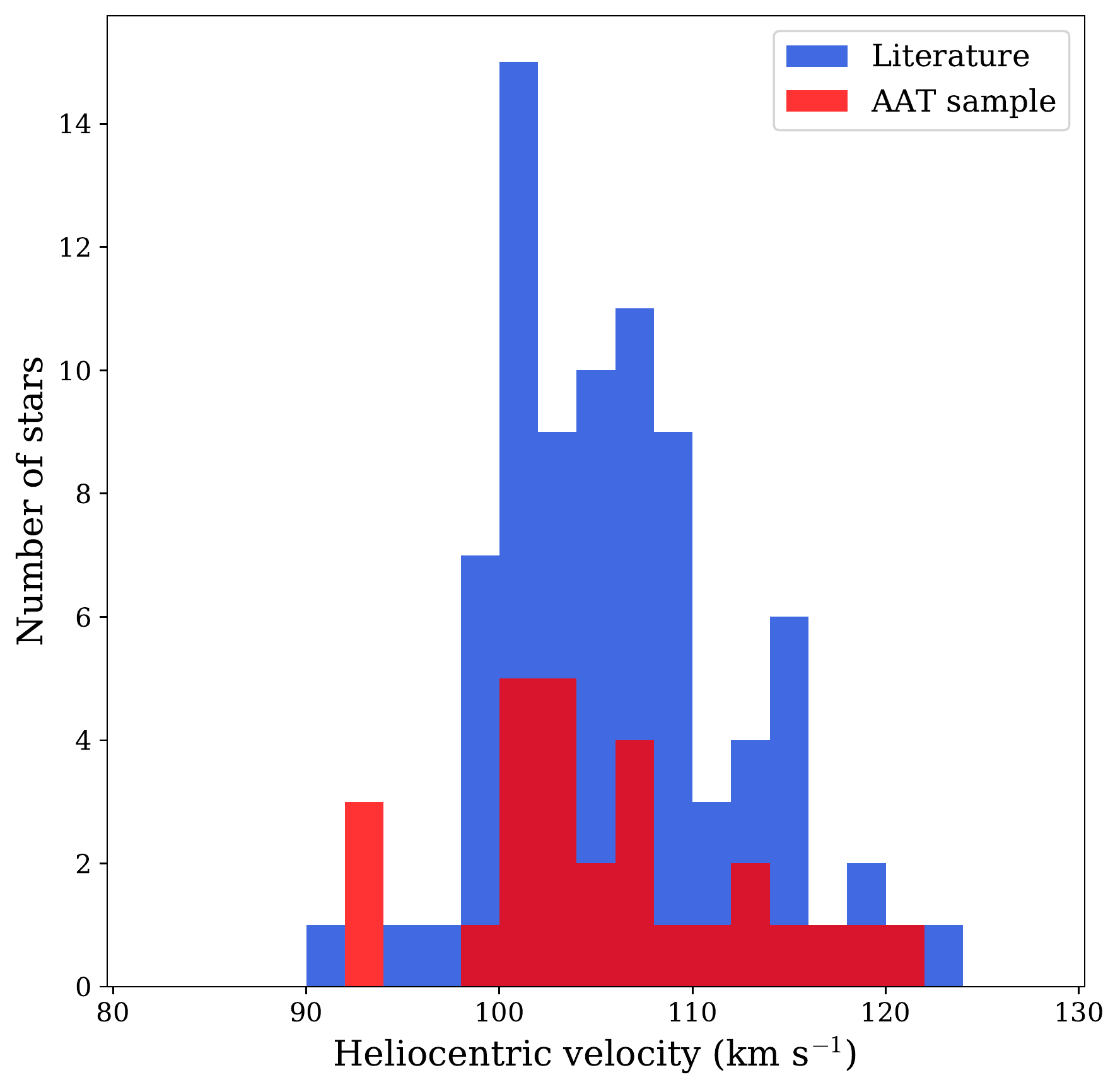}}
\caption{Radial velocity histogram of known Boo~I stars from the literature (blue) and from this work (red).}
\label{hist_vel} 
\end{center}
\end{figure}

\begin{figure}
\begin{center}
\centerline{\includegraphics[width=\hsize]{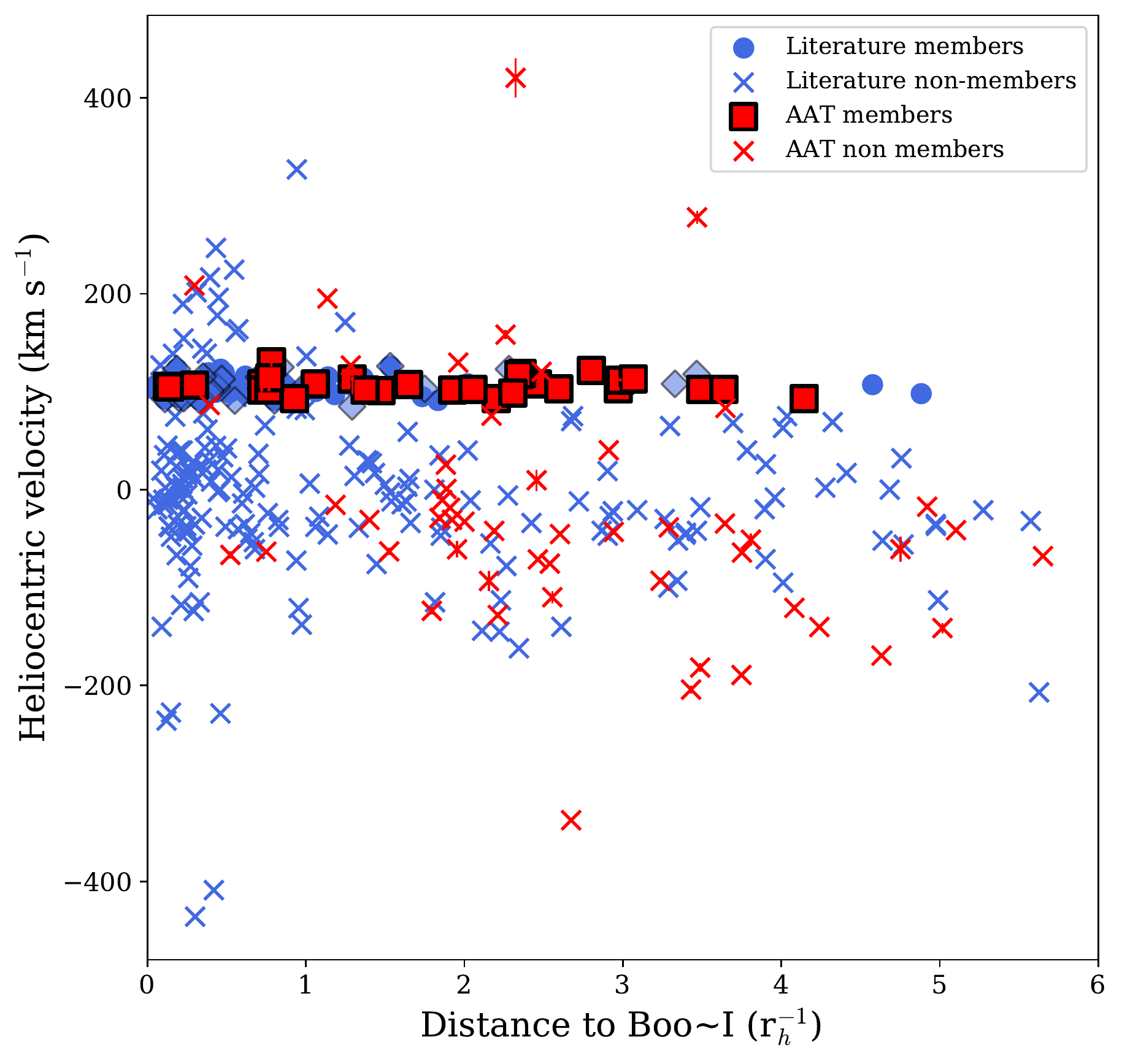}}
\caption{Radial velocities of stars from the literature (blue) and our AAT sample (red) with respect to their distance to Boo~I centroid. Non member of both the literature and the AAT sample are shown as crosses.}
\label{v_vs_r} 
\end{center}
\end{figure}

\begin{figure}
\begin{center}
\centerline{\includegraphics[width=\hsize]{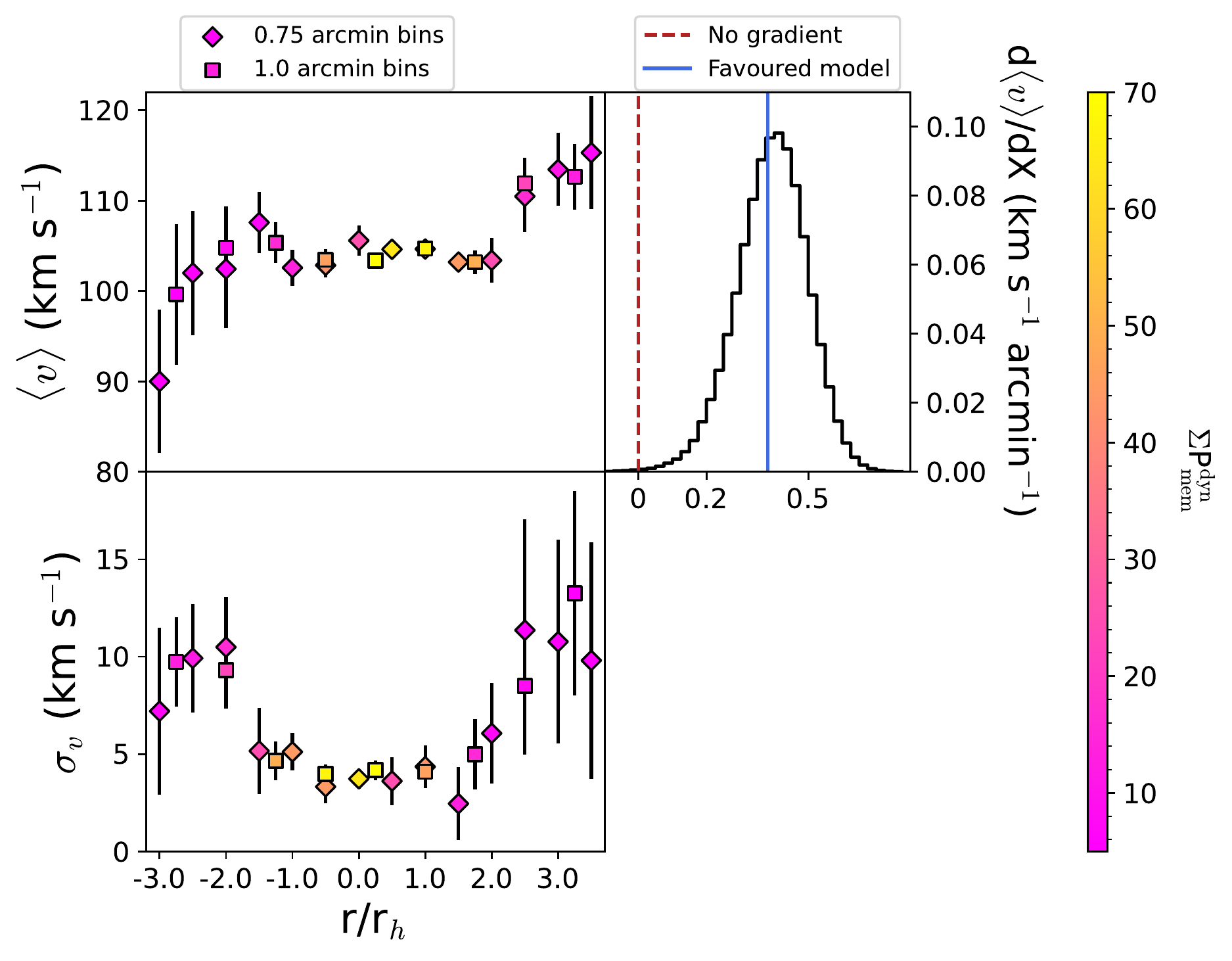}}
\caption{Systemic velocity (\textit{top left panel}) and velocity dispersion (\textit{bottom left panel}) as a function of the distance along the semi-major axis of Boo~I, scaled by the half-light radius. Negative distances are defined as pointing towards negative declinations. Diamonds and squares respectively represent the 0.75 and 1.0 arcmin bins cases. Each point is colour-coded according to the sum of the membership probability of all stars in a given bin. The \textit{top right panel} shows the systemic velocity gradient PDF. The red dashed line indicates the nul gradient and the blue solid line the favoured model.}
\label{gradient_velocity} 
\end{center}
\end{figure}

Using the pipeline described in section 2.2, the velocities of all stars in our sample are derived. Spectra with a S/N ratio of 3 or lower are discarded from our analysis as their resulting velocities are not reliable, leaving our spectroscopic sample with $81$ stars with a median S/N of $\sim 9.1$. To derive reliable velocity uncertainties and uncertainty floor, we apply the following relation of \citet{li19}:

\begin{equation*}
\begin{aligned}
\delta_\mathrm{v} = \sqrt{(1.28\delta^\mathrm{fit}_\mathrm{v})^2 + 0.66^2},
\end{aligned}
\end{equation*}

\noindent with $\delta^\mathrm{fit}_\mathrm{v}$ the velocity uncertainty measured by the MCMC alone, and $\delta_\mathrm{v}$ the final uncertainty considered in the rest of this work. The main properties of the dataset can be found in Table 1. The velocity distribution of the AAT sample is shown in Figure \ref{hist_vel} while the velocities as a function of distance are shown in Figure \ref{v_vs_r}. To assign dynamical membership probability, we derive the radial velocity and velocity dispersion of our sample, through a MCMC algorithm, by maximising the following likelihood: 

\begin{equation}
\begin{aligned}
\mathcal{L}(\langle \mathrm{v}_{\mathrm{BooI}} \rangle, \sigma_{\mathrm{v}}^{\mathrm{1}}, \langle \mathrm{v}_{\mathrm{MW}} \rangle,\sigma_{\mathrm{v}}^{\mathrm{MW}} | \mathrm{v}_{\mathrm{r},k}, \delta_{\mathrm{v},k}) = \\
\Pi_k (\eta_{\mathrm{BooI}} \mathcal{G}(\mathrm{v}_{\mathrm{r},k}, \delta_{\mathrm{v},k} | \langle \mathrm{v}_{\mathrm{1}} \rangle, \sigma_{\mathrm{v}}^{\mathrm{1}}) - \\
 (1 - \eta_{\mathrm{BooI}})\mathcal{G}(\mathrm{v}_{\mathrm{r},k}, \delta_{\mathrm{v},k} | \langle \mathrm{v}_{\mathrm{MW}} \rangle, \sigma_{\mathrm{v}}^{\mathrm{MW}}),
\end{aligned}
\end{equation}\

\noindent with $\eta_\mathrm{BooI}$ the fraction of Boo~I stars and $\sigma^{1}_{\mathrm{v}} = \sqrt{((\sigma_{\mathrm{v}}^{\mathrm{BooI}})^{2} + \delta_{\mathrm{v},k}^{2})}$ where $\delta_{\mathrm{v},k}$ is the individual velocity uncertainty of the $k$-th star and $\sigma_\mathrm{BooI}$ is the intrinsic velocity dispersion of Boo~I (respectively for $\sigma_\mathrm{MW}$ and the MW stars). $\langle \mathrm{v}_{\mathrm{BooI}} \rangle$ and $\langle \mathrm{v}_{\mathrm{MW}} \rangle$ stand for the systemic radial velocity of the system and the MW stars respectively. From this sample alone, we find a systemic radial velocity of $104.7 \pm 1.6$ km s$^{-1}$ and a velocity dispersion of $9.4 \pm 2.0$ km s$^{-1}$. To identify the final number of new member stars found, both the AAT sample and the datasets from the literature are used. The literature datasets are mainly constituted of \citet{martin07}, \citet{norris08}, \citet{norris10}, \citet{koposov11} and \citet{jenkins21}. Some of the stars in these datasets have been re-observed at higher resolution through the years (\citealt{feltzing09}, \citealt{norris10a}, \citealt{lai11}, \citealt{gilmore13}, \citealt{ishigaki14} and \citealt{frebel16}). In that case, the metallicity considered in this work (and velocity when it is also provided) is the one from higher-resolution observations. 

However, most of the past spectroscopic studies were conducted without the PM provided by the Gaia mission. Therefore, some stars identified as members by previous studies are in fact not Boo~I members and are not considered so in the rest of this work. To fold in the Gaia information, the PM membership probability of each star is computed in the same fashion as \citet{longeard20}: we fit a 2D Gaussian mixture model modeling the Boo~I population and the contamination to obtain the systemic PM of the satellite and the local contamination. In order to highlight Boo~I's population among field stars, this procedure is performed on a photometric sample of 1 square degree centered on the satellite for which all stars with a CMD membership probability below 1$\%$ and a Pristine metallicity above $-1.0$ are discarded. The systemic PM obtained is similar to those of \citet{battaglia22} and \citet{mcconnachie20} with $\langle \mu^{*}_\alpha \rangle = -0.39 \pm 0.02$ mas yr$^{-1}$ and $\langle \mu_\delta \rangle = -1.07 \pm 0.01$ mas yr$^{-1}$. The resulting PM membership probabilities are folded in Eq. 3 by scaling the Gaussian distribution representing each population (Boo~I and MW stars) with these probabilities. This re-analysis of the literature shows that 21 stars identified as members by previous studies have too low PM or dynamical probability membership to be considered members of the satellite. Their properties can be found in Table 2. From the entire sample, we find a systemic velocity of $103.0 \pm 0.6$ km s$^{-1}$ and a velocity dispersion of $5.8 \pm 0.5$ km s$^{-1}$, consistent with the re-analysis of Boo~I by \citet{jenkins21}.

Any star with a CMD and dynamical membership probability of respectively at least 10\% and 50\% is considered a new member of Boo~I. The choice for a low CMD probability threshold of 10\% is motivated by the fairly broad distribution of Boo~I stars on its RGB (Figure \ref{field}). We find $27$ spectroscopic members. Only two of these members have a dynamical membership probability below 80\% (with a minimum of $60$\%). Their velocities are shown in Figure \ref{hist_vel}. With 36 targets being promising Boo~I stars in the AAT sample, this yields a member identification success rate of $75$\%. Among those, 5 were identified by \citet{jenkins21} shortly after our AAT data were taken and analysed.  \\

One important aspect of this new sample is that we identified 17 stars beyond 1.0r$_h$ of Boo~I thanks to the 1 degree FoV of the AAT, therefore almost doubling the number of members known this far from the center of the dwarf. Our furthest member is at a distance of 4.1r$_h$.  The dynamics in the outskirts of Boo~I can therefore be studied. We re-perform the dynamical analysis with Equation (3) using subsamples from both the AAT sample and datasets from the literature. Stars are grouped by spatial bins from 0 to 3.5r$_h$ with two different bin sizes: 0.75 and 1 arcminutes. The results are presented in Figure \ref{gradient_velocity} and suggest the existence of spatial gradients. To measure it, we use the formalism of \citet{martin_jin11} and add a Gaussian distribution to account for the contamination from the MW stars:

\begin{equation}
\begin{aligned}
\mathcal{L}(\langle \mathrm{v}_{\mathrm{BooI}} \rangle, \langle \mathrm{v}_{\mathrm{MW}} \rangle, \sigma^\mathrm{BooI}_{\mathrm{v}}, \sigma^\mathrm{MW}_{\mathrm{v}}, d\mathrm{v}/d\chi, \theta | \mathrm{v}_{\mathrm{r},k}, \delta_{\mathrm{v},k}) = \\
\Pi_k \;  \eta_\mathrm{BooI} (\frac{1}{\sqrt{2 \pi \sigma}}) \times \mathrm{exp}(\frac{1}{2}\Delta_\mathrm{v}/\sigma^{2}) + \\
(1 - \eta_\mathrm{BooI}) \mathcal{G}(\mathrm{v}_{\mathrm{r},k}, \delta_{\mathrm{v},k},\langle \mathrm{v}_{\mathrm{MW}} \rangle, \sigma^\mathrm{MW}_{\mathrm{v}}),
\end{aligned}
\end{equation}\

\noindent We define $\Delta_\mathrm{v}$ such as $\Delta_\mathrm{v} = \mathrm{v}_{\mathrm{r},k} - y \times d\mathrm{v}/d\chi + \langle \mathrm{v}_{\mathrm{BooI}} \rangle$ with d$\mathrm{v}$/d$\chi$ the systemic radial velocity gradient. $y$ is the angular distance computed such that $y_k = X_k\cos{\theta} + Y_k\sin{\theta}$ and $\theta$ the position angle of the velocity gradient. We also define $\sigma = \sqrt{(\sigma^\mathrm{BooI}_\mathrm{v} + \delta_\mathrm{v}^{2})}$. We emphasize that this model defines the velocity gradient as a ``linear'' change from one end of the galaxy to the other. Finally, since the velocity measurements are taken from different spectroscopic datasets, such a model could be biased by systematic offsets between different spectrographs and observational setups. To take this into account, we follow the formalism of \citet{minor19} that add an offset parameter for each dataset. Since they are unknown, these parameters are free and will also be derived by the MCMC algorithm.

The resulting velocity offsets between the different spectroscopic datasets are shown in Table 3 and show consistent results: datasets with the exact same setups (\citealt{norris08}/\citealt{norris10} and \citealt{koposov11}/\citealt{jenkins21}) have similar offsets. Furthermore, the velocity offset between our AAT sample and \citet{martin07} of $7.2 \pm 1.6$ km s$^{-1}$ is compatible with the one measured between our dataset and the one of \citet{norris08} ($2.8 \pm 1.5$ km s$^{-1}$) being added to the one between \citet{norris08}  and \citet{martin07} as measured by the former ($4.6$ km s$^{-1}$).
We detect a systemic velocity gradient of d$v$/d$\chi$ $= 0.40 \pm 0.10$ km s$^{-1}$ arcmin$^{-1}$. This translates into a $\sim 4.0$ km s$^{-1}$ shift per $r_h$. To investigate whether the introduction of a gradient in the dynamical properties of the satellite impacts the velocity dispersion of Boo~I and therefore its dynamical mass, they are derived without any gradient for all stars from both our sample and the previous spectroscopic studies from the literature. When no gradient is included in the dynamical model, the velocity dispersion is $5.8 \pm 0.5$ km s$^{-1}$ and is larger than when it is allowed to vary with distance ($4.5 \pm 0.3$ km s$^{-1}$). This result is expected since allowing the systemic velocity to vary will naturally explain part of the dispersion of individual velocities found in a system. However, it has an impact on the mass of the satellite ($108$ M$_\odot$ L$^{-1}_\odot$ with vs. $190$ M$_\odot$ L$^{-1}_\odot$ without any gradient). 

The bottom left panel of Figure \ref{gradient_velocity} also shows an increase of the velocity dispersion at the edges of Boo~I which is a behaviour that has already been observed in UDFs (e.g. \citealt{martin16_tri}).

%The bottom left panel of Figure \ref{gradient_velocity} shows a break feature at around $2$r$_h$ for the velocity dispersion. However, it it not found when considering the largest 1.0 arcmin bins that suggest a flat velocity dispersion. It is therefore not clear whether there is indeed a rapid, localised drop in $\sigma_v$ at this distance or if it just an effect caused by the two smaller bin sizes.

\subsection{Metallicity properties}

\begin{figure}
\begin{center}
\centerline{\includegraphics[width=\hsize]{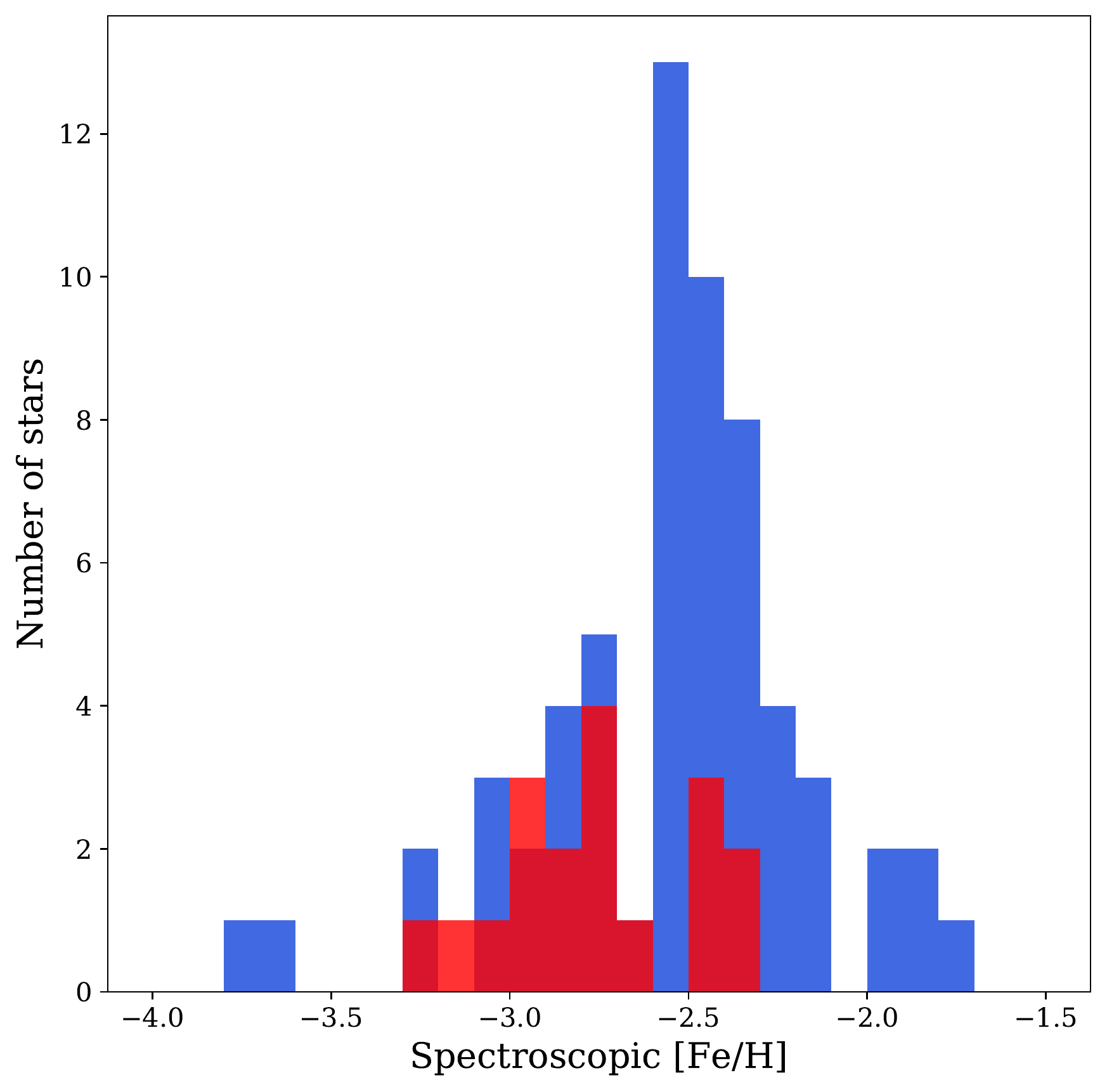}}
\caption{Spectroscopic metallicity histogram of known Boo~I stars from the literature (blue) and from this work (red).}
\label{hist_FeH} 
\end{center}
\end{figure}

\begin{figure}
\begin{center}
\centerline{\includegraphics[width=\hsize]{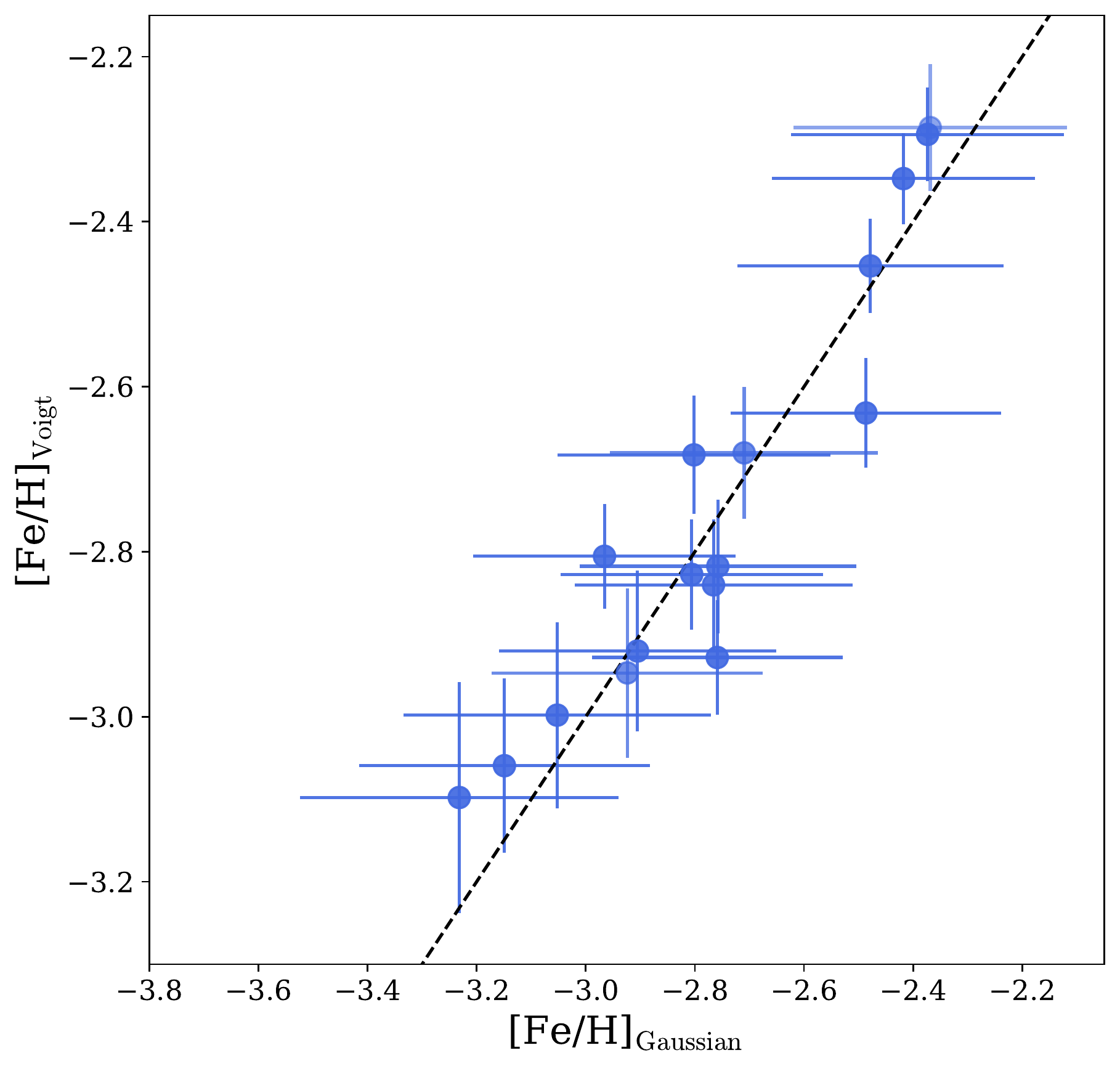}}
\caption{Comparison between the spectroscopic metallicities obtained with Gaussian line profiles and the S10 calibration on the x-axis and Voigt line profiles and the C13 calibration on the y-axis. The circles' opacity represent the dynamical membership probability of each star.}
\label{G_vs_V} 
\end{center}
\end{figure}

\begin{figure}
\begin{center}
\centerline{\includegraphics[width=\hsize]{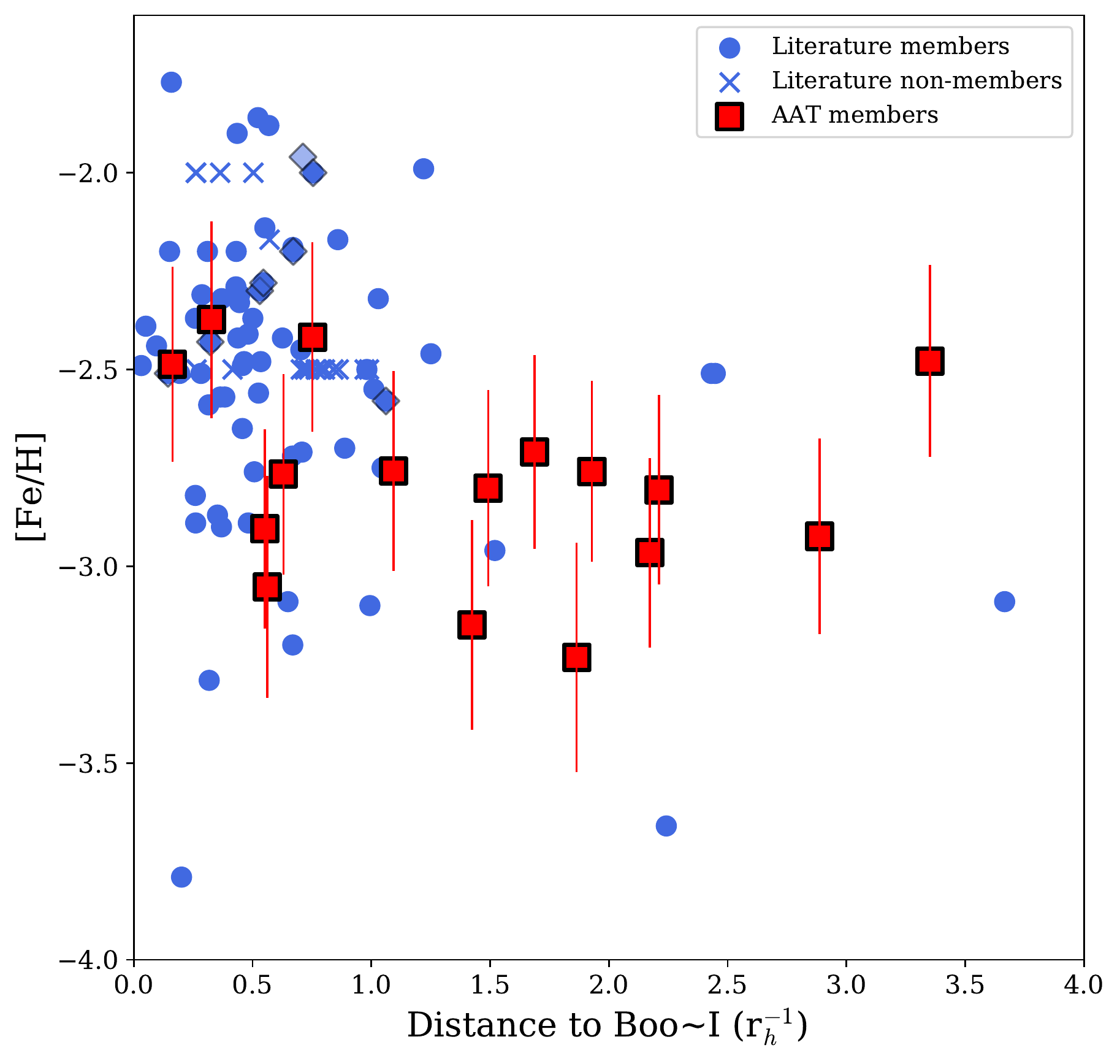}}
\caption{Spectroscopic metallicities of stars from the literature (blue) and our AAT sample (red) with respect to their distance to Boo~I centroid. Blue transparent diamonds show literature members that are not compatible with the dynamical properties of Boo~I found in this work. Crosses shows literature non members.}
\label{FeH_vs_r} 
\end{center}
\end{figure}

\begin{figure}
\begin{center}
\centerline{\includegraphics[width=\hsize]{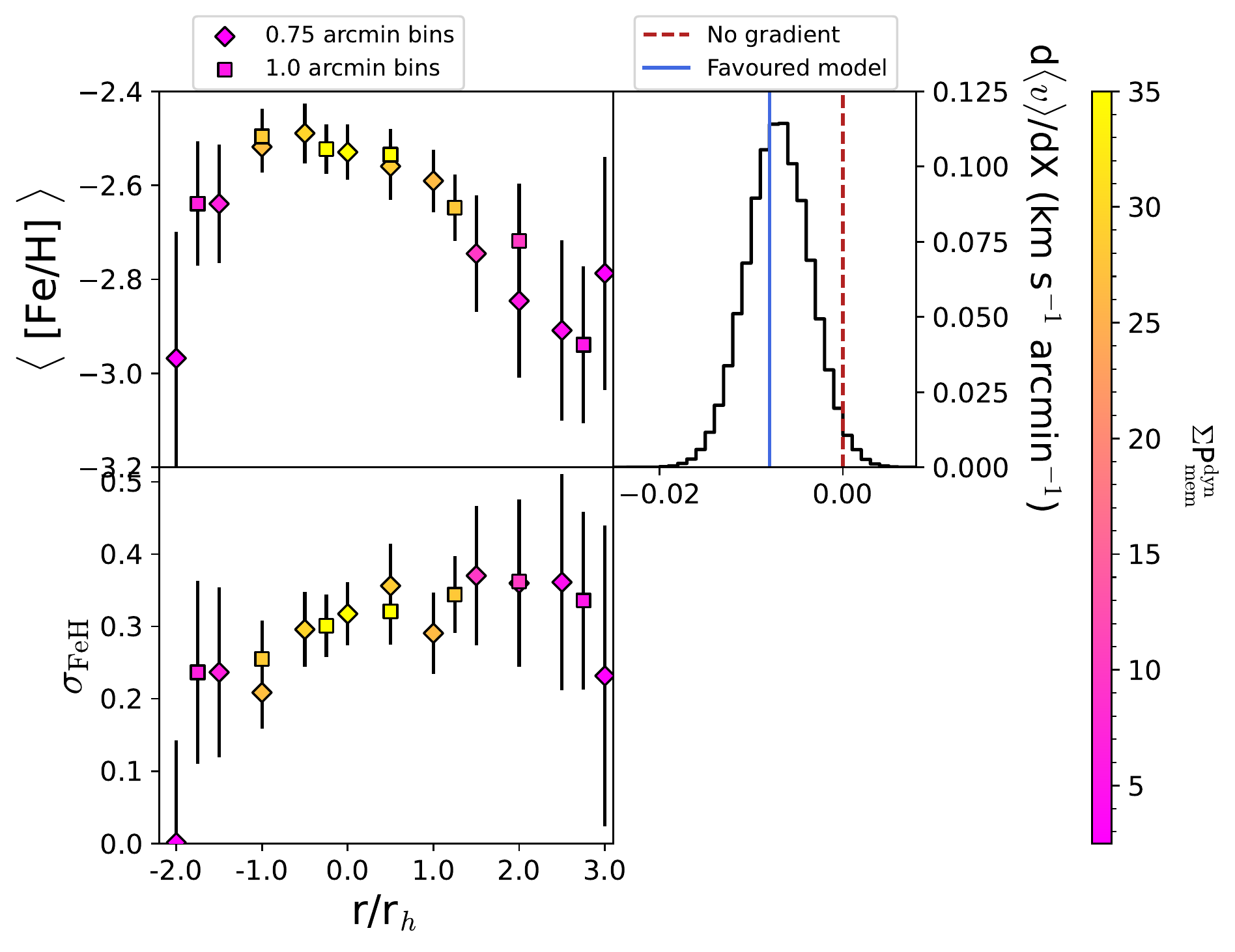}}
\caption{Similar plot as Figure \ref{gradient_velocity} for the metallicity and metallicity dispersion.}
\label{gradient_FeH} 
\end{center}
\end{figure}

Spectroscopic metallicities are derived for the $16$ member stars in our sample with a S/N ratio greater than 10 to ensure reliable EWs measurements. Two different calibrations are used: the one of \citet[S10]{starkenburg10} that applies for Gaussian profiles fit and the one of \citet[C13]{carrera13} for Voigt profiles fit. Both translate the calcium triplet lines EWs into a metallicity measurement reliably down to a metallicity of $ -4.0$. 

To derive their uncertainties, we perform a Monte-Carlo procedure. We randomly draw values of the EWs from their Probability Distribution Functions (PDFs) for 10,000 iterations. At each iteration, we compute the spectroscopic metallicity. The individual photometric and Boo~I distance uncertainties are also folded in in the same way. Finally, to account for the uncertainty on the calibration relation itself, we use an uncertainty of 8\% on each of its coefficient as specified by S10 and the proposed uncertainties on the coefficients stated by C13. For both, these uncertainties are also folded in the Monte Carlo. This process enables us to build a PDF for the metallicity of each star that takes into account the uncertainties on all the parameters involved in the derivation of the spectroscopic metallicities. The 8\% uncertainty of the S10 calibration naturally dominates over all other uncertainty sources, which explains the large uncertainties for the S10 calibration compared to C13.

Figure \ref{hist_FeH} shows the resulting Voigt metallicity distribution of our sample for newly found Boo~I members superimposed with the one in the literature. Figure \ref{G_vs_V} illustrates the comparison between the metallicities obtained with Gaussian and Voigt line profiles and shows that they are perfectly compatible with a negligible bias of $-0.04 \pm 0.07$ dex. Using the equivalent of Eq. 3 for the metallicities, we find a mean metallicity of $-2.7 \pm 0.2$ and a metallicity dispersion of $0.26 \pm 0.05$ from the AAT sample alone. Using the whole literature yields a systemic metallicity of $-2.60 \pm 0.03$ and a metallicity dispersion of $0.34 \pm 0.03$. We find a total of 3 extremely metal-poor stars (EMPs, [Fe/H] $< -3.0$) which accounts for almost half of the total number of EMPs known in the system according to the SAGA database.

The metallicities as a function of distance to Boo~I are shown in Figure \ref{FeH_vs_r} and shows that our sample more than doubles the number of members with known spectroscopic metallicities located further than $1$r$_h$ of the satellite. If the most metal-poor stars of Boo~I are present at all distances, Figure \ref{FeH_vs_r} hints that the more metal-rich population is more centrally concentrated as it almost disappears at a distance above 1.5r$_h$. Such a distribution is expected in higher mass dwarf galaxies in which gas tends to concentrate in the center of the system with time, therefore giving birth to a more centrally-concentrated, metal-rich population (\citealt{leaman13}, \citealt{kacharov17}, \citealt{revaz18} and references therein). To study in more details the potential spatial dependency of Boo~I's metallicity properties, we perform the same analysis as in section 3.1 but for the metallicity. The results are shown in Figure \ref{gradient_FeH}. We detect a small but resolved systemic metallicity gradient of $-0.008 \pm 0.003$ dex arcmin$^{-1}$, translating into a $\sim 0.08$ dex shift per $r_h$. Our systemic metallicity is discrepant from the one of \citet{jenkins21}, who finds a value of $-2.35^{+0.09}_{-0.08}$, at $2.9\sigma$. Two main reasons can be advanced to explain this discrepancy. First, this work re-analyses the entire literature with updated metallicities from subsequent higher-resolution analyses over the years, contrarily to \citet{jenkins21} who analyse their Very Large Telescope (VLT) sample. The second reason is that our sample is constituted of a large number of stars in the outer regions of Boo~I that are more metal-poor than the rest of Boo~I stars because of the existence of a negative metallicity gradient, driving the overall metallicity of our sample lower.

The bottom left panel shows no sign of a metallicity dispersion gradient. Furthermore, similarly to the case of the velocity dispersion, we find that taking a potential gradient into account deflates the metallicity dispersion to $0.26 \pm 0.03$ in the gradient case (vs. $0.34 \pm 0.03$ when no gradient is being measured).

\subsection{A system influenced by tides ?}

\begin{figure*}
\begin{center}
\centerline{\includegraphics[width=0.8\hsize]{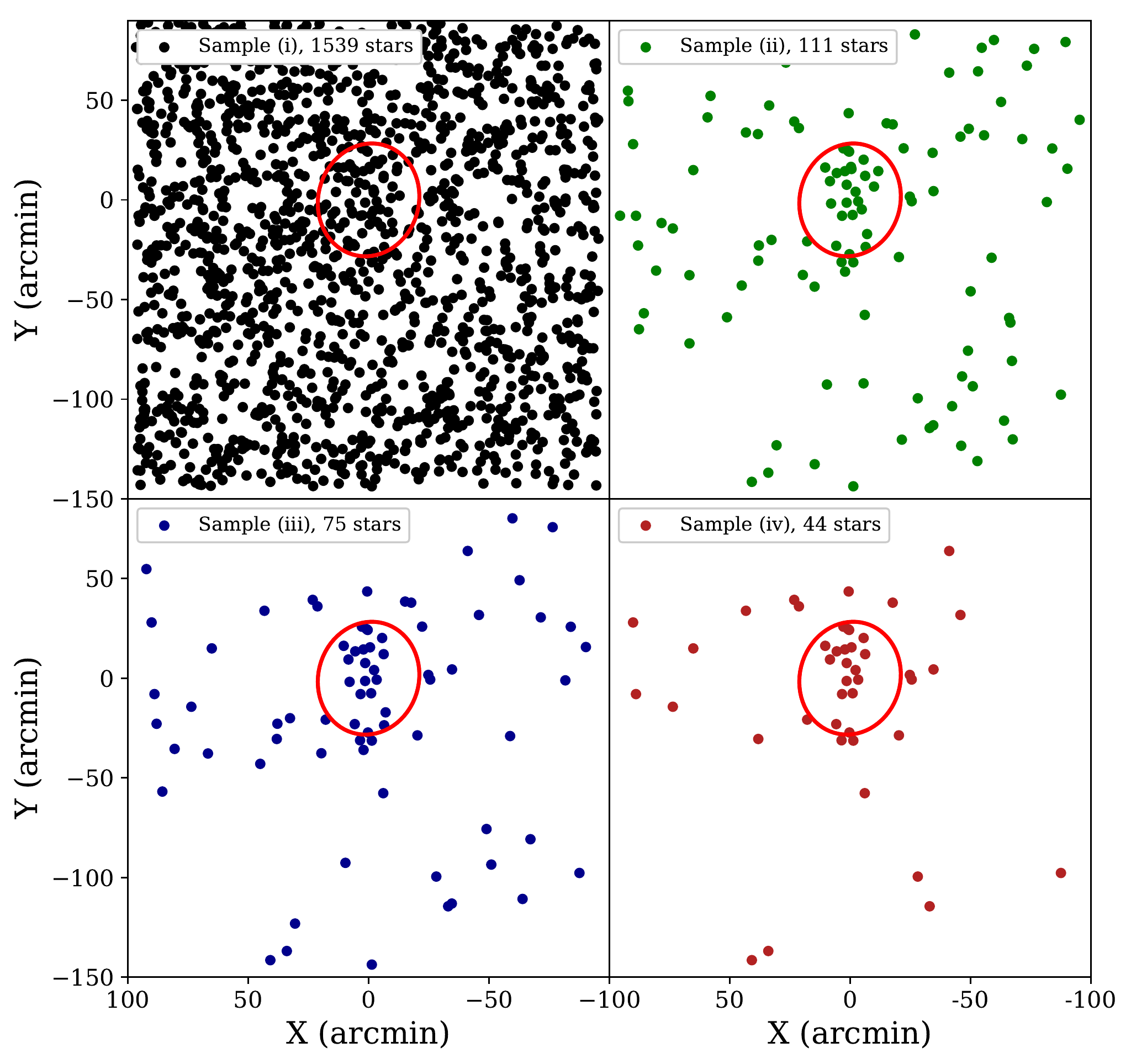}}
\caption{Spatial distributions of stars in the Boo~I field for the four samples detailed in section 3: only photometry (top left), photometry + Gaia (top right), photometry + Gaia + [Fe/H]$_\mathrm{Pristine} < -1.0$ (bottom left) and photometry + Gaia + [Fe/H]$_\mathrm{Pristine} < -2.0$ (bottom right) . The two half-light radii of Boo~I as derived by M18 are shown as the red ellipses.} 
\label{spatial_samples} 
\end{center}
\end{figure*}

\begin{figure*}
\begin{center}
\centerline{\includegraphics[width=0.7\hsize]{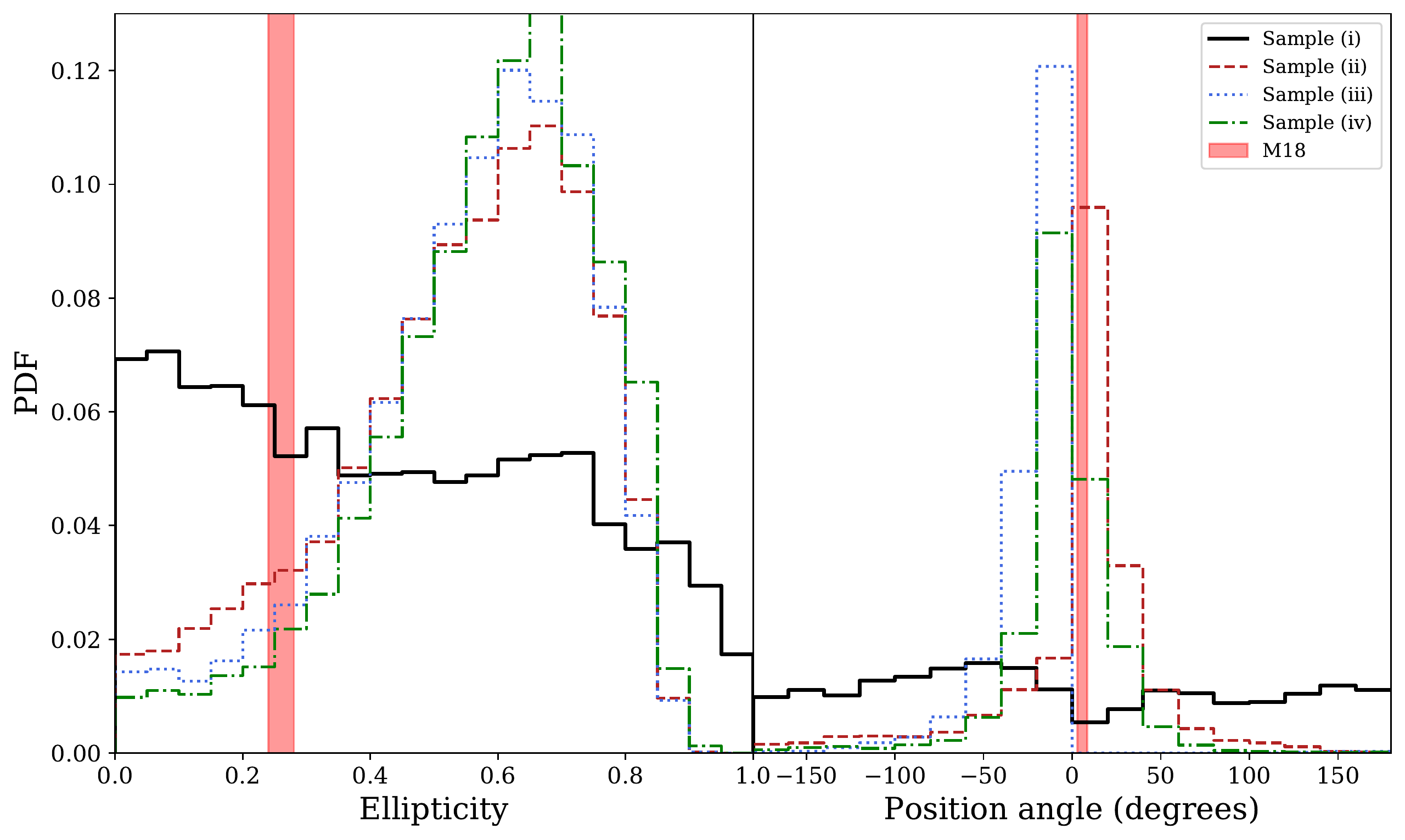}}
\caption{PDFs of the spatial parameters of Boo~I for the four subsamples defined in section 3.3 for the ellipticity (left panel) and the positon angle (right panel). The purely photometric sample (i) is represented by the black solid lines. Adding PMs selection yields the sample (ii) shown as a red dashed line. The sample (iii) characterised by taking sample (ii) and discarding all stars with Pristine metallicities above $-1.0$ is shown as the blue dotted line. Finally, the fourth sample which takes sample (ii) and discards all stars with a Pristine metallicity above $-2.0$ is represented by the dashdotted line. The $1\sigma$ interval of M18 are shown as the red area.} 
\label{spatial_PDFs} 
\end{center}
\end{figure*}

\begin{figure}
\begin{center}
\centerline{\includegraphics[width=0.8\hsize]{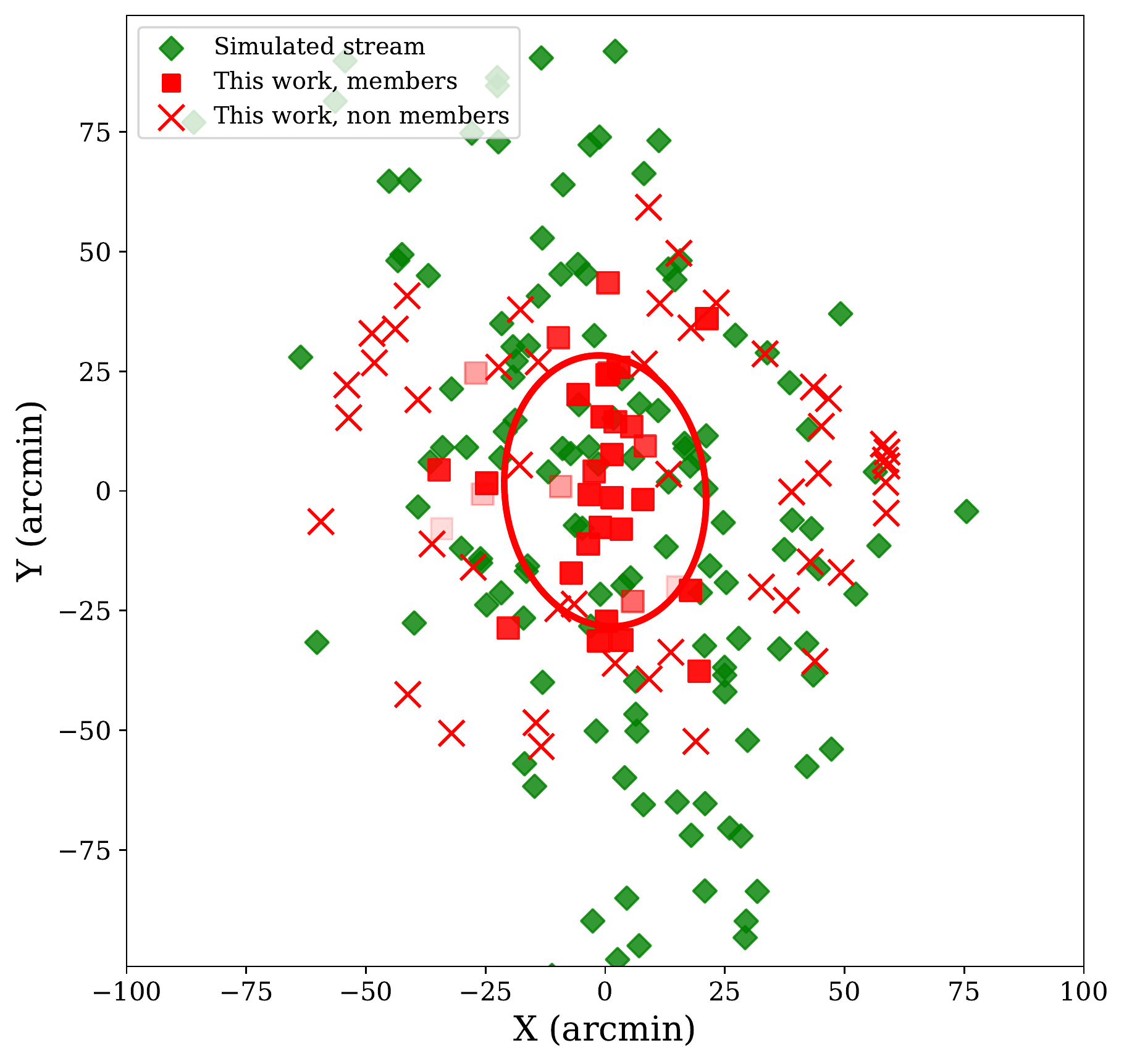}}
\caption{Spatial distribution of test particles in the hypothesis that Boo~I is tidal disrupting (green diamonds) obtained with our toy model described in section 3.3. The spectroscopic sample presented in this work is shown in red crosses (non members) and squares (members). The two half-light radii of Boo~I as derived by M18 are shown with the red ellipse.} 
\label{stream} 
\end{center}
\end{figure}

The detection of a systemic velocity gradient in Boo~I and the spatial distribution of the new members shown in Figure \ref{field} begs the question of the shape of Boo~I and its possible breaking of dynamical equilibrium. Even though the FoV is a two degrees diameter circle, the members appear to be distributed according to a more elongated structure aligned towards the North/South direction. Such a shape is not found by M18, who used deep photometry with a one degree square FoV and concluded that Boo~I's ellipticity is low ($\epsilon_\mathrm{M18} = 0.26 \pm 0.02$) assuming an exponential profile.

To investigate whether the elongated shape of Boo~I detected in our data is significant, we derive the spatial profile of the system for four different cases: 

\begin{enumerate}

\item a SDSS purely photometric sample for which the stars are selected only based on their proximity to Boo~I's RGB with any star with a ($g-i$)$_0$ further than 0.15 of the RGB being discarded.

\item a combination of the sample (i) with the Gaia eDR3. Using the PMs of all stars in the sample and knowing the systemic PM of Boo~I \citep{mcconnachie20}, all stars not compatible within 2$\sigma$ of the system's PM are discarded.

\item a combination of the sample (ii) with the Pristine survey. Using the Pristine photometric metallicities, all stars with a [Fe/H]$_\mathrm{Pristine} > -1.0$ are discarded.

\item a combination of the sample (ii) with the Pristine survey, this time limiting ourselves only to very metal-poor stars ([Fe/H]$_\mathrm{Pristine} < -2.0$), much more representative the systemic metallicity of Boo~I ($\langle$[Fe/H]$_{\mathrm{BooI}}\rangle \sim -2.35$).

\end{enumerate}

To do so, the method of \citet{martin07} is used. It derives spatial parameters assuming an exponential density profile for the satellite and a constant density for field stars. This exponential spatial profile can be written as :

\begin{equation}
\begin{aligned}
\rho_\mathrm{dwarf}(r) = \frac{1.68^2}{2\pi r_h^2(1 - \epsilon)} \mathrm{exp}(-1.68\frac{r}{r_h})
\end{aligned}
\end{equation}

\noindent In Eq. 5, $\epsilon$ is the ellipticity and $r$ the elliptical radius defined as:

\begin{equation}
\begin{aligned}
r = [(\frac{1}{1-\epsilon}((X - X_0)\mathrm{cos}\theta - (Y - Y_0)\mathrm{sin}\theta)^2) + \\
((X - X_0)\mathrm{sin}\theta - (Y - Y_0)\mathrm{cos}\theta)^2)]^{1/2}
\end{aligned}
\end{equation}

\noindent with $X_0$ and $Y_0$ the system's centroid and $\theta$ the position angle.

Since we use a shallower photometric sample than that of M18, the results for the sample (i) will not be exactly the same nor as precise. The goal of this analysis is to assess whether the elongated shape of the members found in this work comes from the addition, with respect to M18, of the Gaia PMs and/or Pristine metallicities, especially at higher distances. 

The spatial distribution of the four samples is shown in Figure \ref{spatial_samples}, which shows that the elongation of Boo~I is not clear with sample (i) using only photometry but becomes obvious as soon as PMs are introduced with the second sample. The shape is conserved when restricting to only very metal-poor stars according to the Pristine survey. To have a more quantitative view, the spatial properties of each sample is derived in the same way as in \citet{martin08}. The resulting PDFs for the ellipticity and position angle are shown in Figure \ref{spatial_PDFs}. For the sample (i), i.e. photometry only, the ellipticty is not resolved. This result is compatible with the deeper data of M18. However, introducing the PMs instantly shows that Boo~I is more elongated that anticipated with a position angle aligned towards the North/South direction ($\epsilon_\mathrm{i} = 0.68 \pm 0.15$, $\theta_\mathrm{i} = 6 \pm 24^{\circ}$). This result is confirmed when the photometric metallicities are used to further discriminate Boo~I stellar population. Furthermore, a recent study of Blue Horizontal Branch (BHB) stars in the vicinity of Boo~I also finds an elongated structure in the same direction as this work \citep{filion21}.

To investigate whether this elongation could be caused by tidal interactions with the MW, we design a toy model that aims to study the spatial distribution that a tidally-disrupting Boo~I should have. The model has been constructed using the particle-spraying method implemented in the {\sc Gala} package \citep{price_whelan20}, with a potential of the MW modelled with the {\sc MWPotential14} of \citet{bovy15} and the adopted distribution function for the stream model of \citet{fardal15}. The auto-gravity is taking into account where the Boo~I is modelled  by a Plummer sphere of mass $4.9 \times 10^6$ M$_\odot$ and with a scale radius of 0.19 kpc. In that configuration 2 particles are sprayed each 0.005 Myr (1 at each Lagrange point). The resulting simulation is shown in Figure \ref{stream} and shows that if Boo~I has been tidally affected by the MW, it should be so in the North/South direction, similar to the direction found in this work using samples (ii), (iii) and (iv). The pericenter of our toy model is quite low ($\sim 34$ kpc) and similar to the pericenter in the heavy MW model of \citet{battaglia22} that find a pericenter of $33.70^{+8.78}_{-7.45}$ kpc. This toy model is not quantitatively perfect but shows that there is indeed an alignement between our spectroscopic members and potential tidal tails in Boo~I.

Finally, our spectroscopic data do not show any trend for the PMs of members with respect to their spatial locations. However, such a trend is not necessarily expected in this case since our furthermost member is located at 4.1r$_h$.

\section{Conclusion}

We observed the Boo~I UFD with new mid-resolution spectroscopy using the AAT and the 2dF multi-object spectrograph to study the outskirts of the satellite. 92 stars were observed including 36 likely Boo~I members based on a PM and Pristine photometric metallicity pre-selection. This pre-selection allows a significantly more successful targeting of members in the outskirts of the system with a success rate of $75$\% at finding new Boo~I stars.

We devised a new pipeline to extract the radial velocities and EWs of stellar spectra around the CaT region. This pipeline is able to fit the CaT lines with Gaussian and Voigt line profiles that allows to derive metallicities with two independent empirical calibrations, i.e. the ones of S10 and C13 respectively. It has proven to perform very well both in terms of radial velocity and metallicities at all S/N regimes when compared with the sample of L21 and 86 Sextans stars from the DART survey.

By analyzing the dynamical properties of this new sample, 27 spectroscopically confirmed Boo~I stars are found. Thanks to the wide FoV of 2dF, 17 members were identified at a distance larger than 1r$_h$ of Boo~I with the furthermost member at 4.1r$_h$, doubling the number of stars known above 1r$_h$. Among those, 12 are located further than 2r$_h$ (vs. 5 in the literature). The spatial extent of the spectroscopic sample allows us to detect a systemic velocity gradient of $0.40 \pm 0.10$ km s$^{-1}$ arcmin$^{-1}$.
Furthermore, the metallicities of 16 of the new members are derived using our new pipeline and the calibrations of S10 and C13. We detect a metallicity gradient in Boo~I of $-0.008 \pm 0.003$ dex arcmin$^{-1}$. Such a detection is not surprising in dwarf galaxies but has mostly been measured in brighter classical dSph and irregulars galaxies, such as Phoenix \citep{kacharov17} or WLM \citep{leaman13}. These metallicity gradients also appear in simulations (\citealt{benitez_llambay16}, \citealt{revaz18}, \citealt{mercado21}). Such a phenomenon is caused by the fact that the gas from which the stars were formed is more centrally concentrated over time, leading to metal-rich stars being preferentially found in the inner region of the galaxy, while older, more metal-poor stars will be distributed more homogeneously across the system. This has been shown in several classical dwarfs such as Sculptor \citep{deboer12a} and Fornax \citep{deboer12b}. However, it is only the second time that a significant metallicity gradient is measured in a UFD after Tucana~II \citep{chiti21}. Furthermore, the high-resolution spectroscopic analysis of \citet{mashonkina17} shows that its knee in $\alpha$ elements abundance, if confirmed, implies that Boo~I was able to reprocess Supernovae~Ia ejecta, which would bring the satellite closer to massive UFDs and classical dwarf galaxies rather than to very low-mass UFDs.

Finally, despite our circular field of view and no \textit{a priori} shape criterion to identify targets, the spatial distribution of member stars is noticeably elongated. This is in contrast with M18, who report a fairly spherical shape of Boo~I ($\epsilon_\mathrm{M18} = 0.26 \pm 0.02$). We have investigated this by studying the shape of the satellite in four cases: using (i) only photometry; (ii) adding PMs; (iii) discarding all stars of (ii) with a Pristine photometric metallicity above -1.0; and (iv) discard all stars of (ii) with a Pristine photometric metallicity above -2.0. When using only photometry, the analysis yields a shape compatible with previous findings with $\epsilon_\mathrm{i} < 0.3$ at the $1\sigma$ level; adding PMs to discriminate between Boo~I and fields stars yields a very elongated shape with $\epsilon_\mathrm{i} = 0.68 \pm 0.13$ following the position angle $\theta_\mathrm{i} = 6 \pm 24$ towards the North/East direction. This result is confirmed when restricting the sample by using photometric metallicities. We investigate whether tidal interactions are a plausible origin to explain this shape and the spatial dependency of the satellite's dynamical properties. To do so, a toy model of a tidally-disrupting  Boo~I under a realistic MW potential is performed and show that, should Boo~I be sufficiently affected by tides, the latter would be oriented in the same direction as that found for Boo~I in sample (ii), (iii) and (iv).

Combining the elongation of the satellite, the detection of both a systemic velocity and metallicity gradient, the $\alpha$ elements abundance from previous studies and the low pericenter of the satellite as derived by \citet{battaglia22} suggests that Boo~I could have been more massive that it is today and that the satellite is currently significantly affected by tidal interactions with the MW.

\section*{Data availability}

The data underlying this article are available in the article.

\section*{Acknowledgments}
This work has been carried out  thanks to the support of  the Swiss National Science Foundation.

AA and NFM acknowledge funding from the European Research Council (ERC) under the European Unions Horizon 2020 research and innovation programme (grant agreement No. 834148).

ES acknowledges funding through VIDI grant "Pushing Galactic Archaeology to its limits" (with project number VI.Vidi.193.093) which is funded by the Dutch Research Council (NWO).

NFM and ZY gratefully acknowledges support from the French National Research Agency (ANR) funded project Pristine (ANR-18-CE31-0017) along with funding from CNRS/INSU through the Programme National Galaxies et Cosmologie and through the CNRS grant PICS07708.

Based on observations obtained with MegaPrime/MegaCam, a joint project of CFHT and CEA/DAPNIA, at the Canada-France-Hawaii Telescope (CFHT) which is operated by the National Research Council (NRC) of Canada, the Institut National des Science de l'Univers of the Centre National de la Recherche Scientifique (CNRS) of France, and the University of Hawaii.

The authors thank the International Space Science Institute, Bern, Switzerland for providing financial support and meeting facilities to the international team Pristine.

This work has made use of data from the European Space Agency (ESA) mission Gaia (https://www.cosmos. esa.int/gaia), processed by the Gaia Data Processing and Analysis Consortium (DPAC, https://www.cosmos.esa. int/web/gaia/dpac/consortium). Funding for the DPAC has been provided by national institutions, in particular the institutions participating in the Gaia Multilateral Agreement.

The Pan-STARRS1 Surveys (PS1) and the PS1 public science archive have been made possible through contributions by the Institute for Astronomy, the University of Hawaii, the Pan-STARRS Project Office, the Max-Planck Society and its participating institutes, the Max Planck Institute for Astronomy, Heidelberg and the Max Planck Institute for Extraterrestrial Physics, Garching, The Johns Hopkins University, Durham University, the University of Edinburgh, the Queen's University Belfast, the Harvard-Smithsonian Center for Astrophysics, the Las Cumbres Observatory Global Telescope Network Incorporated, the National Central University of Taiwan, the Space Telescope Science Institute, the National Aeronautics and Space Administration under Grant No. NNX08AR22G issued through the Planetary Science Division of the NASA Science Mission Directorate, the National Science Foundation Grant No. AST-1238877, the University of Maryland, Eotvos Lorand University (ELTE), the Los Alamos National Laboratory, and the Gordon and Betty Moore Foundation.

Based (in part) on data obtained at Siding Spring Observatory (via program A/2020A/11). We acknowledge the traditional custodians of the land on which the AAT stands, the Gamilaraay people, and pay our respects to elders past and present.

We thank Sven Buder, Sarah Martell and Daniel Zucker who observed the AAT data used in this work.

\clearpage

\newpage

\begin{table*}

\caption{Properties of the new AAT spectroscopic sample. The individual spectroscopic metallicities for Voigt and Gaussian line profiles are reported for stars with S/N $\geq 10$.
\label{table1}}

\setlength{\tabcolsep}{2.5pt}
\renewcommand{\arraystretch}{0.3}
\begin{sideways}
\begin{tabular}{cccccccccccccc}
\hline
RA (deg) & DEC (deg) & $g_0$ & $i_0$ & $CaHK_0$ & $\mathrm{v}_{r} (\kms)$ & $\mu_{\alpha}^{*}$ (mas.yr$^{-1}$) & $\mu_{\delta}$ (mas.yr$^{-1}$) &  S/N & [Fe/H]$^\mathrm{G}_\mathrm{spectro} $ & [Fe/H]$^\mathrm{V}_\mathrm{spectro}$ & [Fe/H]$_\mathrm{Pristine}$ & Member\\

\hline

209.95887 & 14.3290 & 20.2$\pm$0.0 & 20.2$\pm$0.0 & 20.9$\pm$0.1 & 107.7$\pm$2.7 & $-$0.04$\pm$0.44 & $-$2.13$\pm$0.37 & 9.5 & --- & --- & $-$2.27$\pm$0.30 &  Y  \\ \\ 
209.96212 & 14.50078 & 17.4$\pm$0.0 & 17.4$\pm$0.0 & 18.9$\pm$0.0 & 105.7$\pm$0.7 & $-$0.54$\pm$0.06 & $-$1.03$\pm$0.05 & 64.4 & $-$2.4$\pm$0.3 & $-$2.3$\pm$0.1 & $-$2.17$\pm$0.30 &  Y  \\ \\ 
209.98054 & 14.58133 & 19.7$\pm$0.0 & 19.7$\pm$0.0 & 20.5$\pm$0.1 & 85.0$\pm$0.7 & 0.03$\pm$0.35 & $-$1.35$\pm$0.26 & 12.8 & $-$3.5$\pm$0.3 & $-$3.5$\pm$0.2 & $-$2.14$\pm$0.30 &  N  \\ \\ 
210.04425 & 14.63994 & 19.9$\pm$0.0 & 19.9$\pm$0.0 & 20.6$\pm$0.1 & 107.5$\pm$1.9 & 0.11$\pm$0.32 & $-$1.19$\pm$0.24 & 12.0 & $-$3.1$\pm$0.3 & $-$3.0$\pm$0.1 & $-$2.42$\pm$0.30 &  Y  \\ \\ 
210.04458 & 14.49011 & 19.6$\pm$0.0 & 19.6$\pm$0.0 & 20.4$\pm$0.0 & 104.6$\pm$1.4 & $-$0.07$\pm$0.32 & $-$0.99$\pm$0.22 & 12.5 & $-$2.5$\pm$0.2 & $-$2.6$\pm$0.1 & $-$2.28$\pm$0.30 &  Y  \\ \\ 
210.15579 & 14.48278 & 17.4$\pm$0.0 & 17.4$\pm$0.0 & 18.8$\pm$0.0 & 106.5$\pm$0.7 & $-$0.50$\pm$0.06 & $-$1.12$\pm$0.04 & 85.2 & $-$2.4$\pm$0.3 & $-$2.4$\pm$0.1 & $-$1.22$\pm$0.30 &  Y  \\ \\ 
211.031 & 14.43628 & 17.9$\pm$0.0 & 17.9$\pm$0.0 & 18.7$\pm$0.0 & $-$29.2$\pm$3.6 & $-$18.5$\pm$0.10 & $-$24.46$\pm$0.10 & 10.8 & --- & --- & $-$0.95$\pm$0.30 &  N  \\ \\ 
210.59508 & 14.99089 & 19.3$\pm$0.0 & 19.3$\pm$0.0 & 19.8$\pm$0.0 & 40.3$\pm$2.5 & $-$1.25$\pm$0.26 & $-$0.38$\pm$0.22 & 8.9 & --- & --- & $-$2.64$\pm$0.30 &  N  \\ \\ 
209.43221 & 14.38194 & 19.1$\pm$0.0 & 19.1$\pm$0.0 & 19.7$\pm$0.0 & 126.8$\pm$1.4 & $-$0.58$\pm$0.22 & $-$1.70$\pm$0.16 & 16.4 & $-$2.4$\pm$0.2 & $-$2.3$\pm$0.1 & $-$2.34$\pm$0.30 &  N  \\ \\ 
210.06792 & 14.94406 & 19.8$\pm$0.0 & 19.8$\pm$0.0 & 20.4$\pm$0.1 & 104.8$\pm$4.2 & $-$0.36$\pm$0.32 & $-$0.93$\pm$0.23 & 10.5 & $-$3.3$\pm$0.3 & $-$3.1$\pm$0.2 & $-$2.56$\pm$0.30 &  Y  \\ \\ 
210.34625 & 13.64131 & 19.7$\pm$0.0 & 19.7$\pm$0.0 & 20.3$\pm$0.0 & $-$141.2$\pm$5.3 & $-$1.22$\pm$0.36 & $-$2.43$\pm$0.31 & 4.8 & --- & --- & $-$1.77$\pm$0.30 &  N  \\ \\ 
210.11896 & 14.12933 & 19.8$\pm$0.0 & 19.8$\pm$0.0 & 20.5$\pm$0.1 & 93.1$\pm$1.4 & $-$0.43$\pm$0.29 & $-$0.83$\pm$0.29 & 11.4 & $-$2.7$\pm$0.3 & $-$2.7$\pm$0.1 & $-$2.67$\pm$0.30 &  Y  \\ \\ 
210.17817 & 13.85875 & 17.4$\pm$0.0 & 17.4$\pm$0.0 & 18.4$\pm$0.0 & $-$64.2$\pm$0.8 & $-$0.31$\pm$0.08 & $-$1.77$\pm$0.07 & 42.2 & --- & --- & $-$1.78$\pm$0.30 &  N  \\ \\ 
209.63704 & 14.94464 & 19.9$\pm$0.0 & 19.9$\pm$0.0 & 20.7$\pm$0.1 & $-$110.7$\pm$6.2 & --- & --- & 6.4 & --- & --- & $-$1.57$\pm$0.30 &  N  \\ \\ 
211.02067 & 14.67658 & 18.5$\pm$0.0 & 18.5$\pm$0.0 & 19.9$\pm$0.0 & $-$32.5$\pm$1.5 & $-$9.94$\pm$0.14 & $-$8.07$\pm$0.09 & 26.5 & --- & --- & $-$0.85$\pm$0.30 &  N  \\ \\ 
209.84958 & 14.10322 & 20.3$\pm$0.0 & 20.3$\pm$0.0 & 20.7$\pm$0.1 & 123.7$\pm$1.7 & $-$0.41$\pm$0.66 & $-$0.85$\pm$0.66 & 5.0 & --- & --- & $-$1.57$\pm$0.30 &  N  \\ \\ 
209.90879 & 14.11947 & 20.5$\pm$0.0 & 20.5$\pm$0.0 & 21.2$\pm$0.1 & 161.3$\pm$4.4 & $-$1.93$\pm$0.58 & $-$2.08$\pm$0.50 & 6.8 & --- & --- & $-$1.61$\pm$0.30 &  N  \\ \\ 
210.24858 & 14.57167 & 20.1$\pm$0.0 & 20.1$\pm$0.0 & 20.5$\pm$0.1 & $-$66.7$\pm$3.6 & 0.23$\pm$0.52 & $-$1.07$\pm$0.37 & 4.8 & --- & --- & $-$2.39$\pm$0.30 &  N  \\ \\ 
211.03362 & 14.64897 & 20.4$\pm$0.0 & 20.4$\pm$0.0 & 21.4$\pm$0.1 & $-$60.2$\pm$8.3 & $-$18.0$\pm$0.45 & $-$11.75$\pm$0.40 & 3.3 & --- & --- & $-$2.05$\pm$0.30 &  N  \\ \\ 
210.77454 & 13.92003 & 19.3$\pm$0.0 & 19.3$\pm$0.0 & 20.5$\pm$0.0 & $-$34.0$\pm$2.9 & $-$1.30$\pm$0.24 & $-$0.89$\pm$0.20 & 7.8 & --- & --- & $-$0.97$\pm$0.30 &  N  \\ \\ 
209.71508 & 15.14492 & 20.7$\pm$0.0 & 20.7$\pm$0.0 & 21.2$\pm$0.1 & 82.6$\pm$1.9 & 0.01$\pm$0.72 & $-$1.35$\pm$0.65 & 5.1 & --- & --- & $-$4.04$\pm$0.30 &  N  \\ \\ 
209.18975 & 14.95969 & 19.0$\pm$0.0 & 19.0$\pm$0.0 & 19.8$\pm$0.0 & $-$43.3$\pm$3.0 & $-$1.13$\pm$0.18 & $-$0.26$\pm$0.14 & 8.5 & --- & --- & $-$1.52$\pm$0.30 &  N  \\ \\ 
209.57917 & 14.50217 & 19.9$\pm$0.0 & 19.9$\pm$0.0 & 20.6$\pm$0.1 & 130.4$\pm$4.9 & $-$0.14$\pm$0.37 & $-$1.43$\pm$0.32 & 6.5 & --- & --- & $-$2.43$\pm$0.30 &  Y ?  \\ \\ 
210.02442 & 14.05914 & 18.9$\pm$0.0 & 18.9$\pm$0.0 & 19.8$\pm$0.0 & 102.9$\pm$1.1 & $-$0.38$\pm$0.19 & $-$0.88$\pm$0.14 & 23.1 & $-$2.8$\pm$0.2 & $-$2.9$\pm$0.1 & $-$2.17$\pm$0.30 &  Y  \\ \\ 
210.32854 & 15.08133 & 19.8$\pm$0.0 & 19.8$\pm$0.0 & 20.9$\pm$0.1 & $-$38.7$\pm$3.7 & $-$1.47$\pm$0.33 & $-$1.64$\pm$0.23 & 10.5 & --- & --- & $-$1.03$\pm$0.30 &  N  \\ \\ 
209.42333 & 14.58664 & 20.3$\pm$0.0 & 20.3$\pm$0.0 & 21.0$\pm$0.1 & 194.4$\pm$4.8 & $-$0.28$\pm$0.46 & $-$1.57$\pm$0.36 & 7.8 & --- & --- & $-$1.92$\pm$0.30 &  N  \\ \\ 
209.92312 & 14.84953 & 19.2$\pm$0.0 & 19.2$\pm$0.0 & 19.9$\pm$0.0 & 101.6$\pm$1.7 & $-$0.49$\pm$0.21 & $-$1.18$\pm$0.17 & 14.5 & $-$2.8$\pm$0.2 & $-$2.7$\pm$0.1 & $-$3.05$\pm$0.30 &  Y  \\ \\ 
210.16388 & 14.66975 & 20.7$\pm$0.0 & 20.7$\pm$0.0 & 21.3$\pm$0.1 & 91.4$\pm$6.7 & 0.28$\pm$0.55 & $-$0.77$\pm$0.52 & 6.0 & --- & --- & $-$3.98$\pm$0.31 &  Y  \\ \\ 
210.11529 & 14.73794 & 20.5$\pm$0.0 & 20.5$\pm$0.0 & 21.1$\pm$0.1 & 72.8$\pm$2.0 & $-$0.34$\pm$0.47 & $-$1.09$\pm$0.37 & 5.4 & --- & --- & $-$2.46$\pm$0.30 &  N  \\ \\ 
209.46746 & 13.66953 & 18.8$\pm$0.0 & 18.8$\pm$0.0 & 20.1$\pm$0.0 & $-$18.5$\pm$3.2 & $-$0.84$\pm$0.16 & $-$1.65$\pm$0.11 & 8.3 & --- & --- & $-$0.81$\pm$0.30 &  N  \\ \\ 
210.08096 & 13.99397 & 18.7$\pm$0.0 & 18.7$\pm$0.0 & 19.6$\pm$0.0 & 103.0$\pm$1.0 & $-$0.27$\pm$0.15 & $-$1.21$\pm$0.12 & 29.0 & $-$2.8$\pm$0.2 & $-$2.8$\pm$0.1 & $-$2.91$\pm$0.30 &  Y  \\ \\ 
210.00167 & 14.38736 & 19.7$\pm$0.0 & 19.7$\pm$0.0 & 20.4$\pm$0.0 & 101.8$\pm$2.5 & $-$0.30$\pm$0.31 & $-$1.16$\pm$0.25 & 13.3 & $-$2.9$\pm$0.3 & $-$2.9$\pm$0.1 & $-$2.52$\pm$0.30 &  Y  \\ \\ 
209.99687 & 13.99147 & 19.7$\pm$0.0 & 19.7$\pm$0.0 & 20.4$\pm$0.0 & 112.0$\pm$1.4 & $-$0.57$\pm$0.34 & $-$1.50$\pm$0.25 & 8.6 & --- & --- & $-$2.45$\pm$0.30 &  Y  \\ \\ 
209.85212 & 15.04767 & 20.1$\pm$0.0 & 20.1$\pm$0.0 & 20.5$\pm$0.1 & 113.2$\pm$2.5 & $-$0.48$\pm$0.44 & $-$1.45$\pm$0.35 & 9.3 & --- & --- & $-$2.17$\pm$0.30 &  Y  \\ \\ 
209.34629 & 14.83114 & 20.3$\pm$0.0 & 20.3$\pm$0.0 & 20.8$\pm$0.1 & 75.8$\pm$1.6 & 0.06$\pm$0.78 & $-$1.42$\pm$0.55 & 3.8 & --- & --- & $-$1.98$\pm$0.30 &  N  \\ \\ 
210.17567 & 15.50133 & 18.5$\pm$0.0 & 18.5$\pm$0.0 & 19.4$\pm$0.0 & $-$68.1$\pm$1.3 & $-$0.99$\pm$0.14 & $-$0.38$\pm$0.09 & 17.2 & --- & --- & $-$2.67$\pm$0.30 &  N  \\ \\ 
210.030 & 15.23858 & 20.2$\pm$0.0 & 20.2$\pm$0.0 & 20.7$\pm$0.1 & 93.5$\pm$3.8 & $-$0.21$\pm$0.48 & $-$1.03$\pm$0.38 & 8.4 & --- & --- & $-$3.99$\pm$0.30 &  Y  \\ \\ 
209.26517 & 15.07736 & 19.7$\pm$0.0 & 19.7$\pm$0.0 & 20.4$\pm$0.1 & 37.7$\pm$4.6 & 0.72$\pm$0.33 & $-$2.52$\pm$0.27 & 3.9 & --- & --- & $-$0.79$\pm$0.30 &  N  \\ \\ 
210.25617 & 13.95158 & 20.2$\pm$0.0 & 20.2$\pm$0.0 & 20.6$\pm$0.1 & $-$95.5$\pm$5.5 & 0.50$\pm$0.56 & $-$5.34$\pm$0.53 & 7.6 & --- & --- & $-$2.16$\pm$0.30 &  N  \\ \\ 
211.02821 & 14.62108 & 16.7$\pm$0.0 & 16.7$\pm$0.0 & 18.0$\pm$0.0 & 0.8$\pm$1.1 & $-$7.00$\pm$0.04 & $-$4.59$\pm$0.04 & 40.1 & --- & --- & $-$0.84$\pm$0.30 &  N  \\ \\ 
210.05596 & 13.91322 & 19.5$\pm$0.0 & 19.5$\pm$0.0 & 20.4$\pm$0.0 & $-$204.4$\pm$3.9 & $-$0.45$\pm$0.31 & $-$0.78$\pm$0.24 & 6.8 & --- & --- & $-$1.31$\pm$0.30 &  N  \\ \\ 
211.02925 & 14.54336 & 17.7$\pm$0.0 & 17.7$\pm$0.0 & 18.3$\pm$0.0 & $-$124.1$\pm$3.6 & $-$14.91$\pm$0.10 & $-$14.15$\pm$0.09 & 7.2 & --- & --- & $-$1.53$\pm$0.30 &  N  \\ \\ 
209.78967 & 13.62336 & 19.6$\pm$0.0 & 19.6$\pm$0.0 & 20.6$\pm$0.1 & $-$41.6$\pm$2.8 & $-$1.25$\pm$0.26 & $-$0.46$\pm$0.21 & 10.4 & --- & --- & $-$1.38$\pm$0.30 &  N  \\ \\ 
210.16108 & 14.95578 & 19.6$\pm$0.0 & 19.6$\pm$0.0 & 20.7$\pm$0.1 & $-$75.7$\pm$1.6 & $-$1.30$\pm$0.29 & $-$0.76$\pm$0.20 & 13.9 & --- & --- & $-$1.28$\pm$0.30 &  N  \\ \\ 
210.41942 & 15.16842 & 20.2$\pm$0.0 & 20.2$\pm$0.0 & 20.9$\pm$0.1 & $-$51.3$\pm$6.3 & $-$0.93$\pm$0.40 & $-$0.80$\pm$0.31 & 6.0 & --- & --- & $-$2.26$\pm$0.30 &  N  \\ \\ 
210.38567 & 15.11411 & 17.4$\pm$0.0 & 17.4$\pm$0.0 & 18.7$\pm$0.0 & 102.2$\pm$0.7 & $-$0.40$\pm$0.06 & $-$1.02$\pm$0.04 & 78.4 & $-$2.5$\pm$0.2 & $-$2.5$\pm$0.1 & $-$2.46$\pm$0.30 &  Y  \\ \\ 
210.28496 & 15.34111 & 19.9$\pm$0.0 & 19.9$\pm$0.0 & 20.3$\pm$0.0 & $-$23.5$\pm$33.4 & $-$1.60$\pm$0.33 & $-$1.87$\pm$0.26 & 4.8 & --- & --- & $-$2.41$\pm$0.30 &  N  \\ \\ 
210.03404 & 14.92353 & 20.4$\pm$0.0 & 20.4$\pm$0.0 & 21.0$\pm$0.1 & 116.7$\pm$3.5 & 0.43$\pm$0.54 & $-$1.19$\pm$0.41 & 7.3 & --- & --- & $-$2.59$\pm$0.30 &  Y  \\ \\ 
210.32679 & 14.16678 & 18.8$\pm$0.0 & 18.8$\pm$0.0 & 19.5$\pm$0.0 & 102.5$\pm$1.1 & $-$0.15$\pm$0.16 & $-$1.03$\pm$0.11 & 20.1 & $-$3.0$\pm$0.2 & $-$2.8$\pm$0.1 & $-$3.10$\pm$0.30 &  Y  \\ \\ 
210.21658 & 15.16669 & 19.3$\pm$0.0 & 19.3$\pm$0.0 & 19.8$\pm$0.0 & $-$190.9$\pm$5.3 & $-$0.61$\pm$0.29 & 0.26$\pm$0.22 & 10.1 & --- & --- & $-$2.97$\pm$0.30 &  N  \\ \\ 
210.69071 & 14.51019 & 19.4$\pm$0.0 & 19.4$\pm$0.0 & 20.1$\pm$0.0 & $-$15.5$\pm$1.9 & $-$0.82$\pm$0.26 & $-$0.32$\pm$0.19 & 9.5 & --- & --- & $-$1.51$\pm$0.30 &  N  \\ \\ 
208.9975 & 14.40672 & 15.9$\pm$0.0 & 15.9$\pm$0.0 & 16.6$\pm$0.0 & $-$18.2$\pm$1.1 & $-$18.58$\pm$0.04 & 1.60$\pm$0.03 & 37.4 & --- & --- & $-$0.82$\pm$0.30 &  N  \\ \\ 
209.71204 & 14.60381 & 20.0$\pm$0.0 & 20.0$\pm$0.0 & 20.5$\pm$0.1 & $-$63.2$\pm$1.9 & $-$1.90$\pm$0.47 & $-$2.84$\pm$0.38 & 6.1 & --- & --- & $-$1.66$\pm$0.30 &  N  \\ \\ 
210.07733 & 14.38042 & 19.4$\pm$0.0 & 19.4$\pm$0.0 & 20.2$\pm$0.0 & 101.9$\pm$1.5 & $-$0.59$\pm$0.25 & $-$1.14$\pm$0.22 & 17.0 & $-$2.8$\pm$0.2 & $-$2.8$\pm$0.1 & $-$2.45$\pm$0.30 &  Y  \\ \\ 
210.58183 & 14.17881 & 19.8$\pm$0.0 & 19.8$\pm$0.0 & 20.7$\pm$0.1 & $-$92.5$\pm$16.4 & $-$1.28$\pm$0.39 & $-$0.73$\pm$0.28 & 4.1 & --- & --- & $-$1.61$\pm$0.30 &  N  \\ \\

\end{tabular}
\end{sideways}
\end{table*}

\newpage

\begin{table*}
\renewcommand\thetable{1}

\setlength{\tabcolsep}{2.5pt}
\renewcommand{\arraystretch}{0.3}
\begin{sideways}
\begin{tabular}{cccccccccccccc}
RA (deg) & DEC (deg) & $g_0$ & $i_0$ & $CaHK_0$ & $\mathrm{v}_{r} (\kms)$ & $\mu_{\alpha}^{*}$ (mas.yr$^{-1}$) & $\mu_{\delta}$ (mas.yr$^{-1}$) &  S/N & [Fe/H]$^\mathrm{G}_\mathrm{spectro} $ & [Fe/H]$^\mathrm{V}_\mathrm{spectro}$ & [Fe/H]$_\mathrm{Pristine}$ & Member\\

\hline

209.55475 & 14.92425 & 19.7$\pm$0.0 & 19.7$\pm$0.0 & 20.2$\pm$0.0 & 120.6$\pm$2.1 & $-$3.14$\pm$0.35 & $-$2.86$\pm$0.30 & 6.1 & --- & --- & $-$1.62$\pm$0.30 &  N  \\ \\ 
210.02658 & 14.91694 & 19.9$\pm$0.0 & 19.9$\pm$0.0 & 20.6$\pm$0.1 & 98.2$\pm$2.3 & $-$1.15$\pm$0.40 & $-$0.94$\pm$0.27 & 9.5 & --- & --- & $-$2.30$\pm$0.30 &  Y  \\ \\ 
210.79762 & 14.73847 & 19.0$\pm$0.0 & 19.0$\pm$0.0 & 19.8$\pm$0.0 & 25.4$\pm$2.1 & $-$1.85$\pm$0.21 & $-$1.41$\pm$0.17 & 16.1 & --- & --- & $-$1.70$\pm$0.30 &  N  \\ \\ 
210.00958 & 14.77153 & 19.6$\pm$0.0 & 19.6$\pm$0.0 & 20.2$\pm$0.0 & 100.8$\pm$1.7 & $-$0.13$\pm$0.25 & $-$0.95$\pm$0.20 & 15.6 & $-$2.7$\pm$0.3 & $-$2.8$\pm$0.1 & $-$2.96$\pm$0.30 &  Y  \\ \\ 
209.59275 & 14.54056 & 19.4$\pm$0.0 & 19.4$\pm$0.0 & 20.1$\pm$0.0 & 114.1$\pm$0.8 & $-$0.25$\pm$0.25 & $-$0.80$\pm$0.26 & 8.9 & --- & --- & $-$2.30$\pm$0.30 &  Y  \\ \\ 
209.09058 & 14.88239 & 20.3$\pm$0.0 & 20.3$\pm$0.0 & 20.7$\pm$0.1 & $-$338.5$\pm$1.4 & $-$0.54$\pm$0.44 & $-$0.13$\pm$0.36 & 6.2 & --- & --- & $-$1.79$\pm$0.30 &  N  \\ \\ 
209.89825 & 14.22769 & 20.0$\pm$0.0 & 20.0$\pm$0.0 & 21.0$\pm$0.1 & 107.7$\pm$1.9 & $-$0.14$\pm$0.36 & $-$0.73$\pm$0.31 & 10.6 & $-$3.1$\pm$0.3 & $-$3.1$\pm$0.1 & $-$1.25$\pm$0.30 &  Y  \\ \\ 
210.86858 & 14.22953 & 20.1$\pm$0.0 & 20.1$\pm$0.0 & 20.7$\pm$0.1 & $-$127.8$\pm$2.6 & 0.14$\pm$0.41 & 0.29$\pm$0.30 & 4.2 & --- & --- & $-$2.41$\pm$0.30 &  N  \\ \\ 
210.76871 & 14.87539 & 19.3$\pm$0.0 & 19.3$\pm$0.0 & 19.8$\pm$0.0 & 8.5$\pm$9.8 & $-$1.67$\pm$0.24 & $-$3.13$\pm$0.22 & 8.7 & --- & --- & $-$2.82$\pm$0.30 &  N  \\ \\ 
210.35771 & 13.88569 & 20.3$\pm$0.0 & 20.3$\pm$0.0 & 21.0$\pm$0.1 & 102.4$\pm$6.1 & $-$0.30$\pm$0.54 & $-$1.25$\pm$0.36 & 6.0 & --- & --- & $-$1.90$\pm$0.30 &  Y  \\ \\ 
209.85971 & 14.52906 & 20.4$\pm$0.0 & 20.4$\pm$0.0 & 20.8$\pm$0.1 & 207.9$\pm$2.6 & $-$0.84$\pm$0.56 & 0.95$\pm$0.43 & 7.3 & --- & --- & $-$3.99$\pm$0.30 &  N  \\ \\ 
210.28075 & 14.17983 & 19.6$\pm$0.0 & 19.6$\pm$0.0 & 20.4$\pm$0.0 & 129.6$\pm$2.7 & $-$0.79$\pm$0.25 & $-$1.33$\pm$0.21 & 6.2 & --- & --- & $-$1.63$\pm$0.30 &  N  \\ \\ 
209.39729 & 14.32886 & 19.5$\pm$0.0 & 19.5$\pm$0.0 & 20.1$\pm$0.0 & $-$63.4$\pm$3.9 & $-$0.75$\pm$0.27 & $-$0.45$\pm$0.18 & 11.8 & --- & --- & $-$1.97$\pm$0.30 &  N  \\ \\ 
209.09729 & 14.76853 & 19.5$\pm$0.0 & 19.5$\pm$0.0 & 20.4$\pm$0.1 & $-$42.3$\pm$3.0 & $-$1.64$\pm$0.28 & $-$0.54$\pm$0.19 & 4.7 & --- & --- & $-$1.63$\pm$0.30 &  N  \\ \\ 
209.18112 & 15.06208 & 20.0$\pm$0.0 & 20.0$\pm$0.0 & 20.5$\pm$0.1 & 276.7$\pm$5.2 & $-$0.91$\pm$0.40 & $-$1.79$\pm$0.32 & 4.0 & --- & --- & $-$2.42$\pm$0.30 &  N  \\ \\ 
209.77921 & 14.9630 & 20.0$\pm$0.0 & 20.0$\pm$0.0 & 20.6$\pm$0.1 & $-$45.2$\pm$2.9 & $-$0.74$\pm$0.43 & $-$2.30$\pm$0.34 & 6.9 & --- & --- & $-$1.50$\pm$0.30 &  N  \\ \\ 
210.67208 & 14.1325 & 17.8$\pm$0.0 & 17.8$\pm$0.0 & 19.0$\pm$0.0 & $-$71.3$\pm$0.7 & $-$5.99$\pm$0.09 & $-$74.59$\pm$0.06 & 25.1 & --- & --- & $-$1.65$\pm$0.30 &  N  \\ \\ 
209.30996 & 13.80478 & 18.9$\pm$0.0 & 18.9$\pm$0.0 & 19.1$\pm$0.0 & $-$139.4$\pm$0.8 & 0.17$\pm$0.19 & $-$0.10$\pm$0.13 & 9.9 & --- & --- & $-$4.01$\pm$0.30 &  N  \\ \\ 
210.78671 & 14.57481 & 15.5$\pm$0.0 & 15.5$\pm$0.0 & 16.2$\pm$0.0 & $-$31.1$\pm$0.8 & $-$0.54$\pm$0.03 & $-$1.05$\pm$0.03 & 90.6 & --- & --- & $-$0.91$\pm$0.30 &  N  \\ \\ 
209.3075 & 15.19308 & 19.3$\pm$0.0 & 19.3$\pm$0.0 & 20.1$\pm$0.0 & $-$120.6$\pm$3.3 & $-$0.21$\pm$0.26 & $-$0.88$\pm$0.21 & 8.7 & --- & --- & $-$1.30$\pm$0.30 &  N  \\ \\ 
210.82312 & 14.83511 & 20.1$\pm$0.0 & 20.1$\pm$0.0 & 20.5$\pm$0.1 & 399.6$\pm$4.1 & 0.03$\pm$0.37 & $-$0.34$\pm$0.28 & 4.8 & --- & --- & --- &  N  \\ \\ 
210.75746 & 14.26647 & 17.9$\pm$0.0 & 17.9$\pm$0.0 & 18.6$\pm$0.0 & $-$30.9$\pm$2.0 & 0.60$\pm$0.12 & $-$1.68$\pm$0.08 & 17.0 & --- & --- & $-$1.11$\pm$0.30 &  N  \\ \\ 
209.77117 & 13.70644 & 18.3$\pm$0.0 & 18.3$\pm$0.0 & 20.0$\pm$0.0 & $-$169.5$\pm$1.7 & $-$86.19$\pm$0.10 & $-$17.98$\pm$0.08 & 18.4 & --- & --- & $-$1.84$\pm$0.30 &  N  \\ \\ 
211.03329 & 14.60031 & 16.9$\pm$0.0 & 16.9$\pm$0.0 & 17.8$\pm$0.0 & $-$10.5$\pm$1.3 & 3.98$\pm$0.05 & $-$10.19$\pm$0.04 & 32.7 & --- & --- & $-$1.07$\pm$0.30 &  N  \\ \\ 
209.67075 & 14.03533 & 19.2$\pm$0.0 & 19.2$\pm$0.0 & 19.9$\pm$0.0 & 121.6$\pm$1.0 & $-$0.18$\pm$0.20 & $-$1.22$\pm$0.16 & 13.8 & $-$2.9$\pm$0.3 & $-$3.0$\pm$0.1 & $-$3.24$\pm$0.30 &  Y  \\ \\ 
210.05696 & 14.75375 & 20.4$\pm$0.0 & 20.4$\pm$0.0 & 20.9$\pm$0.1 & 102.2$\pm$2.8 & $-$0.58$\pm$0.48 & $-$1.20$\pm$0.36 & 8.1 & --- & --- & $-$3.97$\pm$0.30 &  Y  \\ \\

\end{tabular}
\end{sideways}
\end{table*}

\newpage

\begin{table*}
\caption{List of previous literature members that are found not to be Boo~I members in this work.
\label{table1}}

\setlength{\tabcolsep}{2.5pt}
\renewcommand{\arraystretch}{0.3}
\begin{sideways}
\begin{tabular}{cccccccccccccc}
RA (deg) & DEC (deg) & $\mathrm{v}_{r} (\kms)$ & $\mu_{\alpha}^{*}$ (mas.yr$^{-1}$) & $\mu_{\delta}$ (mas.yr$^{-1}$) &  [Fe/H]$_\mathrm{spectro}$\\

\hline

209.9615 & 14.52061 & 85.2$\pm$5.7 & --- & --- & $-$2.4$\pm$0.2 \\ \\ 
209.91404 & 14.4440 & 97.6$\pm$7.2 & 0.19$\pm$0.47 & $-$1.95$\pm$0.37 & $-$2.2$\pm$0.1 \\ \\ 
209.92579 & 14.49508 & 86.6$\pm$1.8 & $-$0.31$\pm$0.27 & $-$1.47$\pm$0.22 & $-$2.3$\pm$0.1 \\ \\ 
210.01388 & 14.48097 & 115.9$\pm$3.0 & $-$0.97$\pm$0.79 & $-$2.19$\pm$0.64 & $-$2.5$\pm$0.2 \\ \\ 
209.88962 & 14.47267 & 84.4$\pm$6.6 & --- & --- & --- \\ \\ 
209.89325 & 14.50478 & 86.0$\pm$14.8 & --- & --- & --- \\ \\ 
210.02942 & 14.34083 & 105.0$\pm$10.0 & $-$6.67$\pm$0.38 & $-$5.61$\pm$0.33 & --- \\ \\ 
209.94308 & 14.41994 & 91.0$\pm$10.0 & 0.63$\pm$0.48 & $-$2.03$\pm$0.38 & --- \\ \\ 
209.87479 & 14.24908 & 126.0$\pm$10.0 & $-$1.30$\pm$0.58 & $-$1.04$\pm$0.51 & --- \\ \\ 
209.78792 & 14.55314 & 113.0$\pm$10.0 & $-$7.83$\pm$0.63 & $-$3.21$\pm$0.60 & --- \\ \\ 
210.96696 & 14.24264 & 123.0$\pm$10.0 & $-$6.71$\pm$0.18 & $-$8.76$\pm$0.15 & --- \\ \\ 
210.76388 & 13.95244 & 118.0$\pm$10.0 & $-$20.14$\pm$0.28 & $-$4.07$\pm$0.22 & --- \\ \\ 
210.43917 & 14.69853 & 85.0$\pm$10.0 & $-$4.10$\pm$0.41 & $-$3.29$\pm$0.29 & --- \\ \\ 
210.36396 & 14.41219 & 125.0$\pm$10.0 & $-$10.31$\pm$0.49 & 2.93$\pm$0.35 & --- \\ \\ 
209.64854 & 15.08467 & 108.0$\pm$10.0 & $-$12.33$\pm$0.69 & $-$1.68$\pm$0.54 & --- \\ \\ 
209.36458 & 14.28411 & 103.0$\pm$10.0 & $-$1.17$\pm$0.62 & $-$2.26$\pm$0.44 & --- \\ \\ 
209.931 & 14.57731 & 100.5$\pm$1.8 & 0.93$\pm$0.77 & $-$2.38$\pm$0.66 & $-$2.3$\pm$0.2 \\ \\ 
209.93775 & 14.39092 & 102.6$\pm$2.8 & $-$0.73$\pm$2.30 & 2.01$\pm$1.76 & $-$2.0$\pm$0.3 \\ \\ 
209.9405 & 14.37525 & 90.3$\pm$4.1 & --- & --- & $-$2.0$\pm$0.4 \\ \\ 
209.98325 & 14.57378 & 114.7$\pm$0.8 & $-$1.06$\pm$0.31 & $-$2.23$\pm$0.26 & $-$1.2$\pm$0.1 \\ \\ 
210.19812 & 14.40333 & 114.0$\pm$2.1 & 1.30$\pm$0.98 & $-$0.44$\pm$0.77 & $-$2.6$\pm$0.3 \\ \\

\end{tabular}
\end{sideways}
\end{table*}

\begin{table*}

\caption{Velocity offsets of the literature spectroscopic datasets used in this work compared to our AAT sample.
\label{table1}}

\setlength{\tabcolsep}{2.5pt}
\renewcommand{\arraystretch}{0.3}
\begin{tabular}{cccccccccccccc}
\hline
Paper & Setup used &Velocity offset \\
\hline
Martin+07 & Keck/DEIMOS & $7.2 \pm 1.6$ km s$^{-1}$ \\ \\
Norris+08 & AAT/AAOmega (blue spectra) & $2.8 \pm 1.5$ km s$^{-1}$ \\ \\
Norris+10 & AAT/AAOmega (blue spectra) & $2.6 \pm 2.0$ km s$^{-1}$ \\ \\
Koposov+11 & VLT/FLAMES & $0.3 \pm 2.8$ km s$^{-1}$ \\ \\
Jenkins+21 & VLT/FLAMES & $0.3 \pm 1.3$ km s$^{-1}$ \\ \\

\hline

\end{tabular}
\end{table*}

\newcommand{\mnras}{MNRAS}
\newcommand{\pasa}{PASA}
\newcommand{\nat}{Nature}
\newcommand{\araa}{ARAA}
\newcommand{\aj}{AJ}
\newcommand{\apj}{ApJ}
\newcommand{\apjl}{ApJ}
\newcommand{\apjs}{ApJSupp}
\newcommand{\aap}{A\&A}
\newcommand{\aaps}{A\&ASupp}
\newcommand{\pasp}{PASP}
\newcommand{\pasj}{PASJ}

%\bibliography{/Users/longeard/Documents/biblio}
%\bibliographystyle{mn2e}

\clearpage

\end{document}